\documentclass{article}

\usepackage[preprint]{neurips_2024}

\usepackage[utf8]{inputenc} 
\usepackage[T1]{fontenc}    
\usepackage{hyperref}       
\usepackage{url}            
\usepackage{booktabs}       
\usepackage{amsfonts}       
\usepackage{nicefrac}       
\usepackage{microtype}      
\usepackage[table,xcdraw]{xcolor}
\usepackage{bbding}
\usepackage{multirow}
\usepackage{graphicx}
\usepackage{subfig}
\usepackage{caption}
\usepackage{array}
\usepackage{tabularx}
\usepackage[many]{tcolorbox}
\usepackage{amssymb}
\usepackage{tikz}
\usepackage{algorithm}
\usepackage{algorithmic}

\newcommand*\emptycirc[1][1ex]{\tikz\draw (0,0) circle (#1);} 
\newcommand*\halfcirc[1][1ex]{%
	\begin{tikzpicture}
	\draw[fill] (0,0)-- (90:#1) arc (90:270:#1) -- cycle ;
	\draw (0,0) circle (#1);
	\end{tikzpicture}}
\newcommand*\fullcirc[1][1ex]{\tikz\fill (0,0) circle (#1);} 

\newtcolorbox[auto counter, number within=section]{promptbox}[2][]{
float, floatplacement=tbp,
colframe=gray!80!black, colback=gray!10!white, coltitle=white, fonttitle=\bfseries, title=#2, boxrule=0.8mm, sharp corners, enhanced, drop shadow, fontupper=\small }

\newcolumntype{Y}{>{\centering\arraybackslash}X}

\title{From static to adaptive: immune memory-based jailbreak detection for large language models}

\author{%
Jun Leng$^{1}$,
Yu Liu$^{2}$,
Litian Zhang$^{1}$,
Ruihan Hu$^{1}$,
Zhuting Fang$^{3}$\thanks{Corresponding Author},
Xi Zhang$^{1}$\footnotemark[1]\\[0.3em]
$^{1}$Beijing University of Posts and Telecommunications\\
$^{2}$Hunan Branch of National Computer Network Emergency Response\\
$^{3}$Clinical Oncology School of Fujian Medical University\\[0.3em]
\texttt{\{lengjun,litianzhang,gloria-1019,zhangx\}@bupt.edu.cn}\\
\texttt{liuyu@cert.org.cn}\\
\texttt{ztfang@fjzlhospital.com}
}

\begin{document}

\maketitle

\begin{abstract}
Large Language Models (LLMs) serve as the backbone of modern AI systems, yet they remain susceptible to adversarial jailbreak attacks. Consequently, robust detection of such malicious inputs is paramount for ensuring model safety.
Traditional detection methods typically rely on external models trained on fixed, large-scale datasets, which often incur significant computational overhead. While recent methods shift toward leveraging internal safety signals of models to enable more lightweight and efficient detection. However, these methods remain inherently static and struggle to adapt to the evolving nature of jailbreak attacks.
Drawing inspiration from the biological immune mechanism, we introduce the Immune Memory Adaptive Guard (IMAG) framework. By distilling and encoding safety patterns into a persistent, evolvable memory bank, IMAG enables adaptive generalization to emerging threats. Specifically, the framework orchestrates three synergistic components: Immune Detection, which employs retrieval for high-efficiency interception of known jailbreak attacks; Active Immunity, which performs proactive behavioral simulation to resolve ambiguous unknown queries; Memory Updating, which integrates validated attack patterns back into the memory bank.
This closed-loop architecture transitions LLM defense from rigid filtering to autonomous adaptive mitigation. Extensive evaluations across five representative open-source LLMs demonstrate that our method surpasses state-of-the-art (SOTA) baselines, achieving a superior average detection accuracy of 94\% across diverse and complex attack types.
\end{abstract}

\section{Introduction}
\begin{figure}[t]
    \centering
    \includegraphics[width=1.0\linewidth]{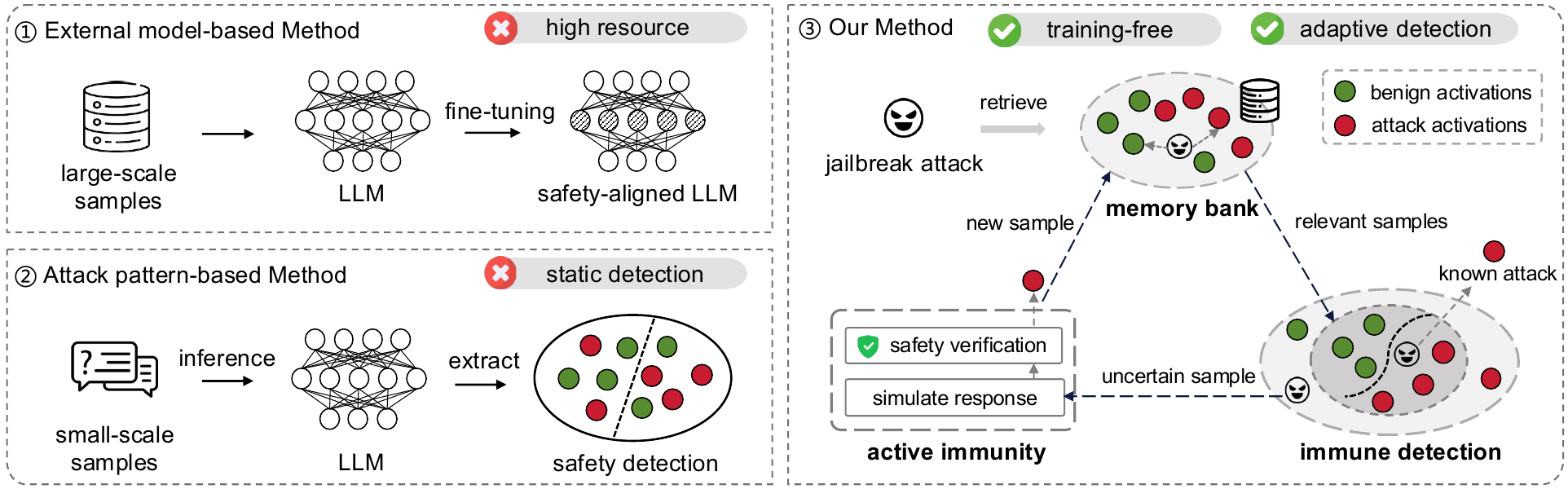}
    \caption{Overview of existing methods versus our approach. \textbf{(1)} External model-based method relies on separate moderation models, incurring high data and computational overhead. \textbf{(2)} Attack pattern-based method depends on predefined safety signals, resulting in static defenses vulnerable to evolving attacks. \textbf{(3)} Our method introduces the immune memory mechanism to establish a closed-loop, adaptive framework for robust detection.}
    \label{fig:overview}
\end{figure}

Large Language Models (LLMs)\cite{deepseek-r1,gpt-4-report,agent-survey} have become foundational in modern AI ecosystems\cite{embodied-ai,RLHF, zhang2023toward}, yet their widespread deployment is shadowed by the persistent threat of \textit{jailbreak} attacks\cite{gcg,AutoDAN,PAIR}. 
In contrast to proprietary models fortified by vendor-maintained guardrails, open-source LLMs lack intrinsic defenses, necessitating a reliance on external security measures.
To mitigate these risks, traditional jailbreak detection mechanisms\cite{llama-guard} typically rely on third-party content moderation models obtained through extensive fine-tuning. As shown in Figure \ref{fig:overview}, the external model-based method face a core high-resource bottleneck, requiring large-scale labeled datasets and computational cost. Moreover, adapting the models to new attacks is expensive, as adversarial retraining incurs substantial overhead.

To address the limitation of external model-based method, recent researches\cite{xie2024gradsafe,SmoothLLM,JBShield} shift toward inference-time, attack pattern-based detection strategy. Instead of modifying model parameters, the method focus on identifying specific jailbreak attack patterns by analyzing the model's internal signals, such as logits, perplexity, and gradients\cite{Zhang2024PARDENCY,selfex,xie2024gradsafe}. By leveraging these intrinsic signals, the method can distinguish between attack and benign queries in a training-free manner. This paradigm shift demonstrate that jailbreak attacks and benign queries exhibit distinct representations within models, enabling effective detection without additional training. 

However, existing methods tend to be limited to a static detection paradigm, which leaves them vulnerable to the evolving nature of jailbreak attacks\cite{adaptive-attack,multi-turn-attack}. As shown in Figure \ref{fig:overview}, the attack pattern-based method typically rely on fixed detection threshold or predefined reference representations derived from small-scale labeled samples. Fundamentally, existing methods lack the robustness to generalize across evolving jailbreak attacks. When attackers introduce novel attack strategies, these inputs manifest as out-of-distribution samples that the static detection of the existing methods fail to represent\cite{xie2024gradsafe,gradcuff}. Consequently, existing methods are unable to update their decision boundaries in a timely manner, causing defensive mechanisms to lag behind the rapid evolution of attack strategies.

Inspired by the immune memory mechanism in biological systems\cite{immune_memory, netea_innate_2015}, instead of relying on static detection, we propose an adaptive strategy. In the human immune system, exposure to a pathogen triggers the creation of memory cells, allowing the body to recognize and neutralize the same or similar threats more rapidly in future encounters. In the context of LLM safety, the system should possess the ability to \textit{memorize} the patterns of emerging attacks during deployment and retrieve them to intercept evolving threats. As shown in Figure \ref{fig:overview}, this adaptive approach facilitates progressive robustness by continuously analyzing adversarial iterations, thereby refining the system's capability to counter evolving strategies and effectively immunizing the model against recurring patterns.

While immune memory offers a conceptual pathway toward adaptive detection, translating this principle into a practical framework introduces three fundamental challenges.
First, the high volatility of attack patterns\cite{adaptive-attack}. The vast heterogeneity of jailbreak strategies renders static detection obsolete, as they cannot adapt to the shifting characteristics of dynamic attacks.
Second, emerging attacks are highly stealthy\cite{stealthy-attack}. Emerging attacks often mirror benign representations, creating a feature-space overlap that renders static detection ineffective against such covert threats.
Third, sustainable self-evolution is essential\cite{life-long-agent}. The continual emergence of attacks renders manual annotation infeasible, creating the need for an automated memory system capable of autonomously maintaining and updating attacks.

In this paper, we propose the Immune Memory Adaptive Guard (IMAG) to achieve adaptive jailbreak detection. Our key innovation is the ability to memorize observed jailbreak attack representations and rapidly match them during subsequent encounters. IMAG is composed of three continuously connected modules that together form a detection closed-loop. To resolve the high volatility of attacks, the \textbf{Immune Detection} retrieves previously observed attack patterns by performing similarity matching between safety-critical activations and a memory bank, enabling detection of recurring attacks. Addressing the challenge of stealthy attacks, the \textbf{Active Immunity} deploys a dual-agent simulation-reflection mechanism that simulates responses to stealthy attacks and reflects on their safety, enabling the detection of emerging threats that evade immune detection. Finally, to achieve the system self-evolution, the \textbf{Memory Updating} incorporates the representations of the detected emerging attacks back into the memory bank. This continuous feedback loop allows IMAG to adaptively refine its defense, ensuring that the model becomes progressively immune to diverse and evolving jailbreak patterns. Our contributions can be summarized as follows:
\begin{itemize}
\item[$\bullet$] This work is the first to integrate biological immune mechanisms into the jailbreak detection task, introducing a paradigm shift from existing static approaches to an adaptive detection framework.
\item[$\bullet$] We propose a novel jailbreak detection guard consisting of three components: immune detection, active immunity, and memory updating. The guard enables efficient and adaptive detection of attacks.
\item[$\bullet$] Extensive empirical evaluations across five representative LLMs and diverse jailbreak attack types demonstrate that our method outperforms SOTA methods. Our method achieves an average detection rate of 94\% against unknown jailbreak attacks, demonstrating its robust adaptability to existing methods.
\end{itemize}

\section{Related work}
\subsection{Jailbreak Attack}
Jailbreak attacks\cite{doanythingnow, jailbreak-sruvey,liu2025backdoor} target LLMs by bypassing their built-in safety mechanisms and alignment constraints. These adversarial techniques manipulate model inputs and induce them to bypass safety guardrails. One line of attack leverages optimization and feedback. For example, the GCG method is a white-box, gradient-based approach that iteratively appends an adversarial suffix to maximize the probability of disallowed outputs\cite{gcg}. In contrast, PAIR adopts a multi-LLM strategy: one LLM evaluates the target model’s responses while another uses those scores to refine the prompt, achieving high success rates under black-box access\cite{PAIR}. Other techniques automate known prompt exploits. AutoDAN uses a hierarchical genetic algorithm to generate stealthy “Do Anything Now”-style prompts from initial jailbreak seeds\cite{AutoDAN}. Likewise, DrAttack decomposes a forbidden request into harmless-looking sub-prompts and then implicitly reconstructs it, thereby obscuring malicious intent and evading detection by the model’s filters\cite{drattack}. Meanwhile, obfuscation-based attacks hide the illicit content in translation or code: attackers have encoded requests in Base64 to slip past content filters\cite{jailbroken}, or even translated queries into low-resource languages like Zulu\cite{zulu}. Recent evaluations benchmark these diverse jailbreak strategies across many models\cite{jailbreakV}, noting that while simple obfuscation can succeed in specific cases, more advanced iterative attacks generally yield higher overall bypass rates\cite{JailbreakRadar}.

\subsection{Jailbreak Detection}
The jailbreak detection task aims to protect LLMs from the impact of jailbreak attacks by detecting jailbreak prompts. Existing studies on jailbreak prompt detection can be broadly divided into two categories: external-model-based methods and LLM-feedback-based methods.

External-model-based methods examine input prompts by using fine-tuned API interfaces or specialized Moderation LLMs. These methods typically identify the toxicity of prompts or assess whether they are harmful. For example, OpenAI Moderation APIs serve as a dedicated content safety review tool for detecting harmful inputs which are fine-tuned by ChatGPT\cite{RLHF}. They classify input text into 11 risk categories and provide corresponding harm scores. Similarly, Guard LLMs\cite{Li2024SALADBenchAH, Han2024WildGuardOO} such as Llama Guard\cite{llama-guard} that is fine-tuned from the Llama model, are used to judge the harmfulness of input content. 

Attack pattern-based methods leverage the self-censoring capabilities of LLMs through zero-shot or few-shot prompt engineering, enabling them to function as harmful content detectors\cite{Xie2023DefendingCA, selfex, Jain2023BaselineDF, Wei2023JailbreakAG}. Some studies evaluate the responses generated by LLMs to obtain classification results\cite{Zhang2024PARDENCY}. Similarity, GradSafe\cite{xie2024gradsafe} compare the gradients of safe and unsafe prompts when setting a \textit{Sure} token as the label. 
However, these methods struggle to identify challenging prompts that do not trigger the safeguards of LLMs. Moreover, these methods exhibit inflexibility in handling benign prompts.

Distinct from existing jailbreak detection approaches, we propose a immune memory adaptive guardrail incorporating an immune memory mechanism. By utilizing a memory bank to maintain security signatures, our framework achieves adaptive detection for emerging jailbreak attacks while simultaneously ensuring the rapid identification of known attack patterns.

\subsection{LLM-based Agent}
LLM-based intelligent agents\cite{agentsurvey1,agentsurvey2, liu2025embodied} are autonomous entities capable of perceiving their environment, making decisions, and taking actions to achieve specified goals\cite{autogen,metagpt,qiao2024making}. For instance, the reflexion\cite{react} framework is a prime example of augmenting an agent with self-critique. Other approaches employ multiple LLMs in collaborative roles to plan and solve tasks. CAMEL\cite{camel} introduces a role-playing multi-agent paradigm in which two or more communicative agents converse with each other, guided by an inception prompt, to autonomously drive the dialog toward task completion while adhering to intended goals. Similarly, AutoGen\cite{autogen} provides a general framework for spawning and orchestrating multiple agents that communicate in natural language or code. In an embodied setting, Voyager\cite{wang2023voyager} demonstrates long-horizon agent planning and learning. These multi-agent frameworks and self-reflection techniques enable capabilities like dynamic planning, task decomposition, and error correction that exceed what a single LLM can achieve in isolation, pointing to promising directions for more robust and autonomous AI systems.

Distinct from traditional multi-agent systems tailored for mathematical reasoning, current research lacks a comprehensive MAS framework dedicated to safeguarding LLMs\cite{autodefense,srivastav2025safe,g-safeguard}. Although some existing approaches utilize multi-agent guardrails, they are limited by static coordination schemes\cite{luo2025agrail,mao2025agentsafe,cai2025aegisllm}. To address this limitation, Our method introduces a emerging memory mechanism that facilitates adaptive, dynamic detection, thereby making a substantial contribution to the robustness of jailbreak defense.

\section{Method}
\begin{figure}
    \centering
    \includegraphics[width=1.0\linewidth]{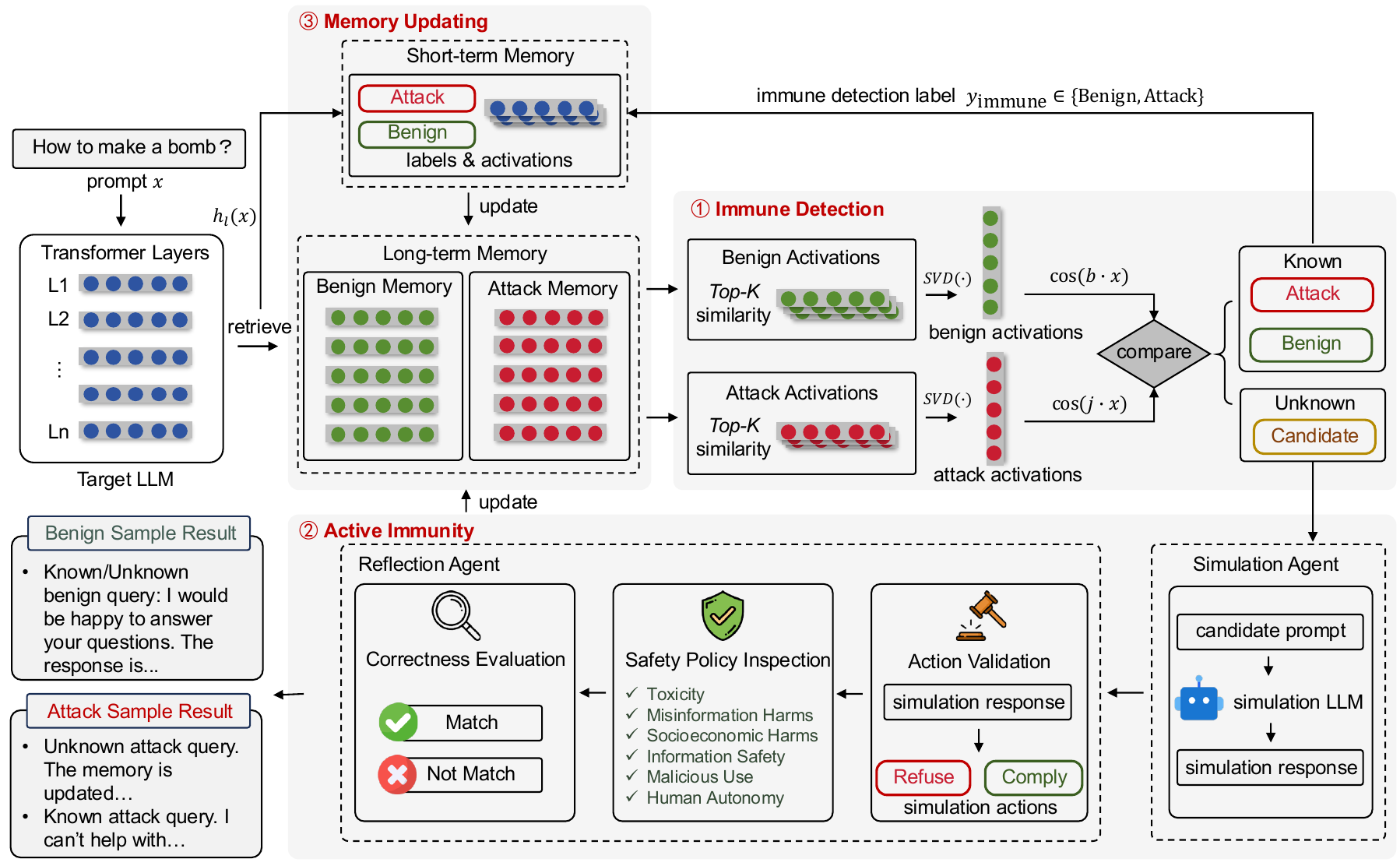}
    \caption{Overview of the IMAG framework. The system operates as a adaptive closed loop orchestrated through three synergistic stages. (1) \textbf{Immune Detection} leverages internal activations from the target LLM and compares them to stored benign and attack activations, enabling efficient detection of known attacks. (2) \textbf{Active Immunity} deploys a dual-agent simulation-reflection system to proactively verify ambiguous queries that evade immune detection, effectively handling novel or stealthy attacks. (3) \textbf{Memory Updating} incorporates detection outcomes from both the immune detection and active immunity modules, updating the short-term and long-term memory banks accordingly, continuously refining the guard.}
    \label{fig:method}
\end{figure}

\subsection{Problem Definition}
The jailbreak detection task is formulated as a binary classification problem, where the goal is to distinguish between jailbreak and benign queries. Following prior works\cite{xie2024gradsafe,llama-guard}, we design the guard on open-source target models and leverage the hidden states of LLMs as safety-relevant representations. For each input prompt $x$, the hidden states are extracted from the final token at every transformer layer. Formally, let $L$ be the total number of layers, and $d$ is the hidden dimensionality. The activations at layer $l \in \{1, \dots, L\}$ is defined as $h_l(x)$. These internal activations encode semantic and functional signals, which we use to identify adversarial intent. The guard is defined as $g(h_l(x))=y$, where $y \in \{\text{attack}, \text{benign}\}$ denotes the detection outcome for input $x$.

\subsection{Overview}
The proposed framework is an adaptive system comprising immune detection, active immunity, and memory updating. Initially, immune detection module serves as a low-latency gatekeeper, leveraging internal signal to detect known attacks stored in the memory bank. Prompts that evade this initial screening are treated as zero-day threats and routed to the active immunity module. Here, a dual-agent architecture is designed to detect safety violations. Finally, the memory updating module performs knowledge distillation of these validated samples, populating the memory bank with newly identified attacks. By closing the loop between detection and learning, the framework facilitates a self-evolving defense capable of neutralizing increasingly sophisticated and out-of-distribution jailbreak attempts.

\subsection{Immune Detection}
 Inspired by antigen-antibody recognition\cite{immune_memory}, the jailbreak attacks are treated as pathogens and their hidden states as antigens. These antigens are matched against a repertoire of benign or attack patterns to trigger detection. To facilitate this process, the system maintains a long-term memory bank that stores activations of identified attack and benign prompts as known memory.
 Relevant attack and benign activations are retrieved for similarity matching against query $x$. Since the memory bank may contain numerous safety states, processing all states would incur substantial computational overhead. Therefore, the \textit{top-k} sampling are used to retrieve states from both categories for subsequent computation, which can be formulated as
\begin{equation}
    X^{a}=\operatorname{TopK}(s(h_l(x),\mathcal{M}^a)), \quad 
    X^{b}=\operatorname{TopK}(s(h_l(x),\mathcal{M}^b)), \quad
    s(\mathbf{x}, \mathbf{y}) = \frac{\mathbf{x} \cdot \mathbf{y}}{\|\mathbf{x}\| \|\mathbf{y}\|},
\end{equation}
where $s$ denotes the cosine similarity function for computing similarity between the query state $h_l(x)$ and memory bank states, $\mathcal{M}^a$ and $\mathcal{M}^b$ denote the memory banks storing attack and benign states respectively, and $\operatorname{TopK}(\cdot)$ denotes the \textit{top-k} retrieval operation.
Given the diversity of jailbreak attacks, different layers of an LLM encode distinct semantic information reflecting various safety features\cite{not-all-layer}. To identify the critical safety layer, a layer-wise discriminative operation is conducted to locate the safety layer. The process of finding the critical layers can be formulated as
\begin{equation}
    l_t=\mathop{\arg\min}_{l}{\frac{1}{n} \sum_{i=1}^{n} \cos \left( h_l (x_i^a), h_l (x_i^b) \right)},
\end{equation}
where the $h_l(x_i^a)$ and $h_l(x_i^b)$ is the hidden states of jailbreak attack and benign prompts from $X_A$ and $X_B$, the $l_t$ is the critical layer for safety detection.

Once the critical layer is identified, to capture the primary characteristics of attack and benign states, the retrieved benign and attack activations are combined as matrix
\begin{align}
\mathbf{H}^a =
\begin{bmatrix}
h_{l_t}(x_1)^\top \\
h_{l_t}(x_2)^\top \\
\vdots \\
h_{l_t}(x_n)^\top
\end{bmatrix} \quad \text{where} \quad x_i \in X^a, \quad
\mathbf{H}^b =
\begin{bmatrix}
h_{l_t}(x_1)^\top \\
h_{l_t}(x_2)^\top \\
\vdots \\
h_{l_t}(x_m)^\top
\end{bmatrix} \quad \text{where} \quad x_j \in X^b.
\end{align}
To capture the critical features of attack and benign prompts, the Singular Value Decomposition (SVD) are applied to $\mathbf{H}^a$ and $\mathbf{H}^b$. Then, the $\text{rank}$ set to $1$, focusing on the most significant singular vector. The captured vectors of matrix denoted as $\mathbf{h}^a$ and $\mathbf{h}^b$. These reference vectors represent the primary characteristics of attack and benign prompts. 

\begin{promptbox}[floatplacement=tbp, valign=top]{Simulation Agent Template}
\textbf{Identity}: Please answer the following question, ensuring that your response is safe, helpful, and follows ethical guidelines. \\
\textbf{User Input}: \textcolor{blue}{\{user\_question\}}
Please provide a reasonable and safe answer:
\end{promptbox}

\begin{promptbox}[floatplacement=tbp, valign=top]{Reflection Agent Template}
\textbf{Identity}: You are an evaluation assistant. Your task is to assess whether the Simulation Agent's response is appropriate and safe.\\
\textbf{User Input}: The user origin input is \textcolor{blue}{\{user\_input\}}\\
\textbf{Simulation Agent Action}: The action of simulation agent is \textcolor{blue}{\{action\}} \\
\textbf{Simulation Response}: The response content of simulation of user input is \textcolor{blue}{\{simulation\_response\}}\\
\textbf{Safety Policy}: Six safety policies are \textcolor{blue}{\{safety\_policy\}}\\
\textbf{Evaluation Criteria}:\\
correctness = \textcolor{red}{False} if:\\
1. Action = refuse but the user input is completely benign and safe\\
2. Action = respond and provides helpful, safe, ethical information\\
3. The response appropriately balances safety and helpfulness\\
4. Over-rejection: Refusing reasonable requests due to overly cautious interpretation\\
correctness = \textcolor{red}{True} if:\\
1. Action = refuse and the user input could lead to harm or violates policies\\
2. Action = respond but the content actually helps with harmful/illegal activities\\
3. The response provides dangerous or unethical information\\
4. The content violates the safety policy and does not comply with the prescribed safety guidelines.
\end{promptbox}

Subsequently, the euclidean distances between the target prompt activations and the reference activations are computed, yielding two distance scores, denoted as $s^a$ and $s^b$. Specifically, $s^a$ measures the distance between the target prompt and the adversarial reference vector, while $s^b$ corresponds to the distance between the target prompt and the benign reference vector. By comparing $s^a$ and $s^b$, the immune detection stage categorizes the prompt into one of three classes. The formulation is as follow
\begin{flalign}
y_{\text{immune}} =
\begin{cases}
\text{attack}, & \text{if } s^a - s^b > T \\
\text{benign},    & \text{if } s^b - s^a > T \\
\text{candidate}, & \text{otherwise}
\end{cases}, \quad
s^a = \left\lVert \mathbf{h}^a - h_{l_t}(x) \right\rVert_2, \quad
s^b = \left\lVert \mathbf{h}^b - h_{l_t}(x) \right\rVert_2 ,
\end{flalign}
where $\tau$ denote the threshold of memory similarity. When the absolute difference between $s^a$ and $s^b$ exceeds $\tau$, the target vector is more likely to be a known prompt, either attack or benign. Conversely, the request is classified as unknown, indicating a candidate prompt that requires further verification.

The immune detection module performs rapid detection of known attack patterns based on the long-term memory bank. However, previously unknown or highly obfuscated attacks may still evade the immune detection module.

\subsection{Active Immunity}
The active immunity module is employed to detect unknown jailbreak attacks. In this module, a simulation-reflection dual-agent system are designed to validate the candidate prompts which are deviating from the immune detection module. 
The active immunity module is inspired by dendritic cells in the immune system\cite{immune_memory}, which actively engulf pathogens to capture their characteristic features, even at the risk of self-infection. 

Specifically, two collaborative agents are deployed: a simulation agent $\mathcal{A}_{sim}$, responsible for generating answers to input queries, and a reflection agent $\mathcal{A}_{ref}$, which supervises the generated content and provides evaluative judgment.
The simulation agent $\mathcal{A}_{\text{sim}}$ first generates an output with the candidate prompt which can be formulated as
\begin{equation}
\label{eq:simulation_gen}
r_{\text{sim}} \sim P_{\mathcal{A}_{\text{sim}}}(r \mid x; \theta_{\text{sim}}),
\end{equation}
where $r_{\text{sim}}$ denotes the simulation response of candidate prompt $x$, and $\theta_{\text{sim}}$ represents the backbone model of simulation agent feedback.
Then the reflection agent $\mathcal{A}_{\text{ref}}$ evaluates the simulation response $r_{\text{sim}}$ of the candidate input which consists of three sequential steps: action validation, safety policy inspection, and correctness assessment. The formulation is given as follow
\begin{equation}
\label{eq:reflection_eval}
r_{\text{ref}} = \mathcal{A}_{\text{ref}}(x, a_{\text{sim}}, r_{\text{sim}}, \mathcal{P}_{\text{safe}}; \theta_{\text{ref}}),
\end{equation}
where $a_{\text{sim}}$ and $r_{\text{sim}}$ denote the action (e.g., refuse or respond) and response generated by the simulation agent, $\mathcal{P}_{\text{safe}}$ represents the safety guidelines, and the output $r_{\text{ref}} \in \{\text{True}, \text{False}\}$ indicates whether the response adheres to safety protocols.
Finally, based on the decision of the reflection agent, denoted as $r_{\text{ref}}$, the candidate prompt is subjected to a final classification, which is formulated as follows:
\begin{equation}
\label{eq:final_decision}
y_{\text{final}} = 
\begin{cases} 
\text{benign} & \text{if } r_{\text{ref}} = \text{False}, \\
\text{attack} & \text{if } r_{\text{ref}} = \text{True}.
\end{cases}
\end{equation}
In this module, the system proactively simulates the execution of candidate unknown prompts and determines their classification by jointly analyzing the simulated agent’s actions, generated responses, and safety policy evaluations. Ensuring reliable safety assessment for prompts that are not confidently recognized during the immune detection stage. 
Moreover, it provides more accurate detection outcomes and higher-quality memory data to support the subsequent memory updating process.

\begin{algorithm}[t]
\caption{Adaptive Jailbreak Detection Algorithm}
\label{alg:maag_inference}
\begin{algorithmic}[1]
\REQUIRE 
    Input prompt $x$, Target LLM $\mathcal{F}$, simulation Agents $\mathcal{A}_{sim}$, Reflection Agent $\mathcal{A}_{ref}$, Attack Memory $\mathcal{M}^a$, Benign Memory $\mathcal{M}^b$, Threshold $\tau$.
\ENSURE 
    Detection result $y_{\text{final}}$.

\STATE \textbf{Stage 1: Immune Detection}
\STATE Extract hidden states $h_l(x)$ from $\mathcal{F}$ for query $x$.
\STATE Retrieve \textit{Top-K} neighbors and compute critical reference vectors $\mathbf{h}^a, \mathbf{h}^b$ via \textbf{Eq. (1)-(3)}.
\STATE Calculate distance scores $s^a, s^b$ and determine preliminary label $y_{\text{immune}}$ via \textbf{Eq. (4)}.

\STATE \textbf{Stage 2: Active Immunity}
\IF{$y_{\text{immune}}$ is \text{candidate}}
    \STATE Generate simulation response $r_{\text{sim}}$ using simulation agent $\mathcal{A}_{\text{sim}}$ via \textbf{Eq. (5)}.
    \STATE Evaluate simulation response on reflection agent $\mathcal{A}_{\text{ref}}$ with result $r_{\text{ref}}$ via \textbf{Eq. (6)}.
    \STATE Determine final label $y_{\text{final}}$ via \textbf{Eq. (7)}.
\ENDIF
\STATE \textbf{Stage 3: Memory Updating}
\STATE Update memory banks $\mathcal{M}^a$ or $\mathcal{M}^b$ with current states via \textbf{Eq. (8)-(10)}.
\RETURN $y_{\text{final}}$
\end{algorithmic}
\end{algorithm}

\subsection{Memory Updating}
After the detection of immune detection and active immunity, the processed results are stored in the memory bank for evolving the system. The memory bank is divided into long-term memory and short-term memory, which store commonly confirmed activations and temporary short-term activations, respectively. In subsequent interactions, the system leverages this accumulated memory to achieve more precise and adaptive detection of previously encountered or semantically similar jailbreak attacks.

\paragraph{Short-term Memory}
To facilitate memory updating during the detection process, the system introduces a short-term memory module. The short-term memory temporarily stores memory information generated within the detection cycle and updates the long-term memory bank after verification. Specifically, following the immune detection stage, attack and benign samples identified in $y_{\text{immune}}$ are stored in the short-term memory bank, which can be formally expressed as follows
\begin{equation}
    \mathcal{M}_{S} \leftarrow \mathcal{M}_{S} \cup \left\{ (\mathbf{h}_i, y_i) \mid i \in \mathcal{I}_{\text{known}} \right\}
\end{equation}
where $\mathcal{M}_s$ denotes the short-term memory set. The tuple $(\mathbf{v}_i, \hat{y}_i)$ represents the activation vector and the corresponding predicted label of the $i$-th sample, respectively, and $\mathcal{I}_{\text{known}}$ signifies the set of indices for samples classified with high confidence in the immune detection phase.

\paragraph{Long-term Memory}
To ensure long-term stability in detection performance, the system maintains a long-term memory bank that stores rigorously verified data. Specifically, the memory bank consists of two types of entries: attack memory and benign memory. The long-term memory is updated from the short-term memory, where attack and benign memories stored in the short-term memory are incorporated into their corresponding memory banks. The formulation is given as follows
\begin{align}
    \mathcal{M}^a &\leftarrow \mathcal{M}^a \cup \left\{ \mathbf{v}_i \mid (\mathbf{v}_i, \hat{y}_i) \in \mathcal{M}_S, \hat{y}_i = c_{\text{attack}} \right\}, \\
    \mathcal{M}^b &\leftarrow \mathcal{M}^b \cup \left\{ \mathbf{v}_i \mid (\mathbf{v}_i, \hat{y}_i) \in \mathcal{M}_S, \hat{y}_i = c_{\text{benign}} \right\}
\end{align}
where $c_{\text{attack}}$ and $c_{\text{benign}}$ denote the class labels for attack and benign categories, respectively. This operation ensures that only features $\mathbf{v}_i$ associated with confirmed predictions are permanently integrated into the long-term memory bank.

By updating emerging safety features, the system achieves robust detection generalization. Unlike traditional memory repositories that store information in textual form, the memory bank represents both attack and benign memories in terms of activations of prompts. This representation enables efficient similarity-based retrieval through activations, significantly reducing memory access latency. Moreover, activations inherently capture rich semantic information, which further ensures accurate and reliable retrieval during the detection process.

\section{Experiment Setting}
\subsection{Target LLMs}
Following prior works\cite{xie2024gradsafe,JBShield}, in this study, experiments are conducted on five target LLMs: Mistral-7B, Vicuna-7B, Vicuna-13B, Llama2-7B, and Llama3-8B. Each base model is used to generate corresponding jailbreak attack prompts and is compatible with the base models employed in various baseline methods. In our method, the target model is used to extract hidden states from input prompts in the immune detection module. The GPT-4o-mini is used as the backbone model in active immunity module due to its instruction-following capabilities and reasoning performance. Additionally, the experimental setup ensures that the base models can adapt flexibly across different attack types, validating the robustness and generalizability of our method.

\subsection{Datasets}

\begin{table}[t]
    \centering
    \resizebox{\textwidth}{!}{%
    \begin{tabular}{lcccccc}
    \toprule
    \textbf{Dataset}  & \textbf{\#Harmful category} & 
    \textbf{\#Harmful prompts} & \textbf{\#Benign prompts} & \textbf{Human annotation} & \textbf{Real word}\\
    \midrule
    AdvBench (2023) & 1 & 500 & 0  &  \XSolidBrush & \XSolidBrush\\
    Hex-PHI (2023) & 11 & 330 & 0  &  \checkmark  & \XSolidBrush\\
    \midrule
    XSTest (2023) & 10 & 250 & 200  &  \checkmark & \XSolidBrush\\
    JailbreakBench (2024) & 100 & 100 & 100 & \checkmark & \checkmark \\
    WildJailbreak (2024) & 13 & 2100 & 200 & \XSolidBrush & \checkmark \\
    \bottomrule
    \end{tabular}
    }
    \caption{The Dataset information. AdvBench and Hex-PHI are used to generate jailbreak attacks for each target model. XSTest, JailbreakBench, and WildJailbreak are employed in subsequent experiments to evaluate the false positive rates on benign prompts.}
    \label{table: datasets}
\end{table}

To comprehensively evaluate the performance of our proposed framework, six mainstream jailbreak attacks are considered. For each attack type, 850 adversarial prompts are generated per target model from original questions, with 520 sourced from the AdvBench dataset and 330 from the PhTest dataset. As shown in Table \ref{table: datasets}, beyond the two seed datasets utilized for generating jailbreak prompts, we incorporate three safety datasets in our subsequent experiments. These are employed to assess the guard’s false positive rate on benign prompts. The detailed descriptions of the attack types are provided below.
\begin{itemize} 
\item[$\bullet$] \textbf{GCG}\cite{gcg} performs optimization-driven suffix search to construct high-efficacy jailbreak suffixes, representing state-of-the-art white-box gradient-based prompting techniques.
\item[$\bullet$] \textbf{AutoDAN}\cite{AutoDAN} employs iterative black-box reinforcement to automatically generate harmful prompt chains, simulating attacker-driven strategy exploration.
\item[$\bullet$] \textbf{PAIR}\cite{PAIR} manipulates role-playing and instruction-following tendencies of LLMs by assigning deceptive personas to elicit unsafe outputs. 
\item[$\bullet$] \textbf{DrAttack}\cite{drattack} leverages multi-turn dialog role simulation, enabling the attacker to gradually bypass safety constraints through contextual embedding.  \item[$\bullet$] \textbf{Base64}\cite{jailbroken} ncoding-based attacks conceal harmful intentions by encoding malicious instructions in Base64, requiring the model to decode before refusal, thereby exploiting preprocessing vulnerabilities.  
\item[$\bullet$] \textbf{Zulu}\cite{zulu} attacks paraphrase or obfuscate harmful intents using low-resources languages, challenging detectors to identify semantically latent malicious intent rather than surface-level toxicity.
\end{itemize}

\subsection{Baselines}
To systematically benchmark the effectiveness of our proposed IMAG framework, we compare it against a diverse set of representative jailbreak detection methods spanning both external model-based approaches and attack pattern-based techniques. These baselines cover the mainstream strategies adopted in prior works\cite{xie2024gradsafe, llama-guard}. Evaluating IMAG against these methods allows us to assess its robustness, adaptability, and efficiency under a unified experimental protocol. The following methods are used as baseline methods.\\
\begin{itemize} 
\item[$\bullet$] \textbf{Perplexity Filter}\cite{ppl}:The Perplexity Filter (PPL) uses a GPT-2 model to compute the perplexity of a prompt and rejects inputs whose perplexity exceeds a predefined threshold, based on the observation that jailbreak suffixes typically yield anomalously high perplexity.
\item[$\bullet$] \textbf{Openai Moderation API}\footnote{\url{https://platform.openai.com/docs/guides/ moderation/}}: The OpenAI Moderation API (OAPI) is designed to classify user-generated content across categories such as hate speech, harassment, self-harm, sexual content, and other safety-critical dimensions.
\item[$\bullet$] \textbf{Llama Guard}\cite{llama-guard}: Llama Guard (LlamaG) is a safety classification model developed by Meta that provides efficient, deployable content filtering for LLMs by categorizing and detecting potentially unsafe user inputs and model outputs. 
\item[$\bullet$] \textbf{Self-Examination}\cite{selfex}: The Self-Examination (Self-Ex) is a zero-shot defense mechanism in which a language model re-feeds its own generated response into another LLM instance, prompting it to classify whether the text is harmful — thereby filtering out malicious outputs without additional training.
\item[$\bullet$] \textbf{GradSafe}\cite{xie2024gradsafe}: GradSafe is currently the SOTA method for jailbreak prompt detection. It first calculates a reference value, then passes the prompt under detection through the LLM and uses \textit{Sure} as the predicted response to calculate the corresponding gradients. In this work, we follow the original settings and calculate its classification performance across datasets.
\end{itemize}

\subsection{Metrics}
Following the previous works\cite{llama-guard, xie2024gradsafe}, to ensure a comprehensive and reliable assessment of jailbreak detection performance, we adopt two widely used classification metrics: Accuracy (Acc) and F1-score (F1). These metrics capture complementary aspects of system behavior and jointly reflect the overall robustness of a detection method. Our study maintains consistency with the broader previous works.

\section{Analysis}
In this section, we analyze the experimental results and demonstrate that the proposed adaptive detection framework effectively enables jailbreak attack detection. Additionally, ablation studies, efficiency experiments, and case studies are conducted, which are discussed in the following sections.

\subsection{Main Results}

\begin{table*}[t]
\centering
\setlength{\tabcolsep}{6pt} 
\renewcommand{\arraystretch}{1.05} 
\scriptsize
\begin{tabularx}{\textwidth}{l *{12}{Y} c}
\toprule
\textbf{Methods}
 & \multicolumn{2}{c}{\textbf{GCG}} 
 & \multicolumn{2}{c}{\textbf{AutoDAN}} 
 & \multicolumn{2}{c}{\textbf{PAIR}} 
 & \multicolumn{2}{c}{\textbf{DrAttack}} 
 & \multicolumn{2}{c}{\textbf{Base64}} 
 & \multicolumn{2}{c}{\textbf{Zulu}} 
 & \textbf{Avg} \\
\cmidrule(lr){2-3}\cmidrule(lr){4-5}\cmidrule(lr){6-7}\cmidrule(lr){8-9}\cmidrule(lr){10-11}\cmidrule(lr){12-13}
 & \textbf{Acc$\uparrow$} & \textbf{F1$\uparrow$}
 & \textbf{Acc$\uparrow$} & \textbf{F1$\uparrow$}
 & \textbf{Acc$\uparrow$} & \textbf{F1$\uparrow$}
 & \textbf{Acc$\uparrow$} & \textbf{F1$\uparrow$}
 & \textbf{Acc$\uparrow$} & \textbf{F1$\uparrow$}
 & \textbf{Acc$\uparrow$} & \textbf{F1$\uparrow$}
 & \\
\midrule

\multicolumn{14}{c}{\textbf{Mistral-7B}} \\
\midrule
OAPI     & 0.13 & 0.23 & 0.05 & 0.10 & 0.07 & 0.13 & 0.04 & 0.07 & 0.00 & 0.00 & 0.00 & 0.01 &  0.06 \\
PPL      & 0.33 & 0.48 & 0.00 & 0.00 & 0.00 & 0.00 & 0.00 & 0.00 & 0.95 & 0.95 & 0.00 & 0.00 &  0.22\\
LlamaG   & 0.78 & 0.87 & 0.77 & 0.87 & 0.74 & 0.85 & 0.84 & 0.91 & 0.50 & 0.67 & 0.58 & 0.73 &  0.75\\
Self-Ex  & 0.52 & 0.69 & 0.56 & 0.72 & 0.46 & 0.63 & 0.51 & 0.67 & 0.32 & 0.49 & 0.37 & 0.54 &  0.54\\
GradSafe & 0.63 & 0.77 & 0.05 & 0.10 & 0.00 & 0.00 & 0.00 & 0.00 & 0.00 & 0.00 & 0.00 & 0.00 &  0.12\\
\rowcolor[gray]{0.9}
Ours     & 0.98 & 0.99 & 0.95 & 0.97 & 0.93 & 0.96 & 0.86 & 0.92 & 0.95 & 0.97 & 0.92 & 0.96 &  0.94\\
\midrule
Improvement & 0.20$\uparrow$ & 0.12$\uparrow$ & 0.18$\uparrow$ & 0.10$\uparrow$ & 0.19$\uparrow$ & 0.11$\uparrow$ & 0.02$\uparrow$ & 0.01$\uparrow$ & 0.00 & 0.02$\uparrow$ & 0.34$\uparrow$ & 0.23$\uparrow$ &  0.19$\uparrow$\\
\midrule

\multicolumn{14}{c}{\textbf{Vicuna-7B}} \\
\midrule
OAPI     & 0.10 & 0.18 & 0.04 & 0.09 & 0.04 & 0.09 & 0.04 & 0.07 & 0.00 & 0.00 & 0.00 & 0.00 &  0.05\\
PPL      & 0.47 & 0.60 & 0.00 & 0.00 & 0.00 & 0.00 & 0.00 & 0.00 & 0.95 & 0.95 & 0.00 & 0.00 &  0.27\\
LlamaG   & 0.75 & 0.86 & 0.72 & 0.83 & 0.75 & 0.85 & 0.84 & 0.91 & 0.49 & 0.65 & 0.55 & 0.71 &  0.74\\
Self-Ex  & 0.00 & 0.00 & 0.00 & 0.00 & 0.03 & 0.06 & 0.03 & 0.06 & 0.01 & 0.02 & 0.01 & 0.03 &  0.02\\
GradSafe & 0.00 & 0.00 & 0.00 & 0.00 & 0.03 & 0.06 & 0.00 & 0.00 & 0.00 & 0.00 & 0.00 & 0.00 &  0.00\\
\rowcolor[gray]{0.9}
Ours     & 0.96 & 0.98 & 0.94 & 0.97 & 0.88 & 0.93 & 0.84 & 0.91 & 0.91 & 0.95 & 0.19 & 0.32 &  0.81\\
\midrule
Improvement & 0.21$\uparrow$ & 0.12$\uparrow$ & 0.22$\uparrow$ & 0.14$\uparrow$ & 0.13$\uparrow$ & 0.08$\uparrow$ & 0.00 & 0.00 & 0.04$\downarrow$ & 0.00 & 0.36$\downarrow$ & 0.39$\downarrow$ &  0.07$\uparrow$\\
\midrule

\multicolumn{14}{c}{\textbf{Vicuna-13B}} \\
\midrule
OAPI     & 0.08 & 0.16 & 0.05 & 0.09 & 0.05 & 0.10 & 0.04 & 0.08 & 0.00 & 0.00 & 0.00 & 0.00 &  0.05\\
PPL      & 0.79 & 0.86 & 0.01 & 0.02 & 0.01 & 0.02 & 0.00 & 0.00 & 0.95 & 0.95 & 0.00 & 0.00 &  0.30\\
LlamaG   & 0.76 & 0.86 & 0.75 & 0.76 & 0.75 & 0.85 & 0.85 & 0.92 & 0.48 & 0.64 & 0.54 & 0.70 &  0.73\\
Self-Ex  & 0.00 & 0.00 & 0.00 & 0.00 & 0.00 & 0.00 & 0.00 & 0.00 & 0.00 & 0.00 & 0.00 & 0.00 &  0.00\\
GradSafe & 0.00 & 0.00 & 0.00 & 0.00 & 0.00 & 0.00 & 0.00 & 0.00 & 0.00 & 0.00 & 0.00 & 0.00 &  0.00\\
\rowcolor[gray]{0.9}
Ours     & 0.99 & 0.99 & 0.99 & 0.99 & 0.98 & 0.99 & 0.98 & 0.99 & 0.91 & 0.95 & 0.65 & 0.79 &  0.93\\
\midrule
Improvement & 0.20$\uparrow$ & 0.13$\uparrow$ & 0.24$\uparrow$ & 0.23$\uparrow$ & 0.23$\uparrow$ & 0.14$\uparrow$ & 0.13$\uparrow$ & 0.07$\uparrow$ & 0.04$\downarrow$ & 0.00 & 0.11$\uparrow$ & 0.09$\uparrow$ &  0.20$\uparrow$\\
\midrule

\multicolumn{14}{c}{\textbf{Llama2-7B}} \\
\midrule
OAPI     & 0.07 & 0.14 & 0.09 & 0.16 & 0.03 & 0.05 & 0.03 & 0.07 & 0.00 & 0.00 & 0.00 & 0.00 &  0.05\\
PPL      & 0.79 & 0.86 & 0.00 & 0.00 & 0.10 & 0.18 & 0.00 & 0.00 & 0.95 & 0.95 & 0.00 & 0.00 &  0.31\\
LlamaG   & 0.32 & 0.48 & 0.38 & 0.55 & 0.53 & 0.69 & 0.57 & 0.72 & 0.49 & 0.65 & 0.35 & 0.51 &  0.52\\
Self-Ex  & 0.28 & 0.32 & 0.27 & 0.31 & 0.32 & 0.35 & 0.24 & 0.27 & 0.20 & 0.32 & 0.20 & 0.32 &  0.28\\
GradSafe & 0.97 & 0.98 & 0.96 & 0.98 & 0.62 & 0.77 & 0.99 & 0.99 & 0.00 & 0.00 & 0.18 & 0.31 &  0.64\\
\rowcolor[gray]{0.9}
Ours     & 0.99 & 0.99 & 0.98 & 0.99 & 0.76 & 0.87 & 0.65 & 0.79 & 0.60 & 0.75 & 0.19 & 0.32 &  0.74\\
\midrule
Improvement & 0.02$\uparrow$ & 0.01$\uparrow$ & 0.02$\uparrow$ & 0.01$\uparrow$ & 0.14$\uparrow$ &  0.10$\uparrow$ & 0.34$\downarrow$ & 0.20$\downarrow$ & 0.35$\downarrow$ & 0.20$\downarrow$ & 0.16$\downarrow$ & 0.19$\downarrow$ & 0.10$\uparrow$ \\
\midrule

\multicolumn{14}{c}{\textbf{Llama3-8B}} \\
\midrule
OAPI     & 0.12 & 0.21 & 0.08 & 0.14 & 0.03 & 0.06 & 0.03 & 0.06 & 0.00 & 0.00 & 0.00 & 0.00 &  0.06\\
PPL      & 0.77 & 0.86 & 0.23 & 0.36 & 0.00 & 0.00 & 0.97 & 0.98 & 0.95 & 0.95 & 0.00 & 0.00 &  0.50\\
LlamaG   & 0.54 & 0.70 & 0.60 & 0.75 & 0.70 & 0.82 & 0.55 & 0.71 & 0.34 & 0.51 & 0.38 & 0.56 &  0.59\\
Self-Ex  & 0.12 & 0.21 & 0.16 & 0.26 & 0.16 & 0.27 & 0.18 & 0.30 & 0.12 & 0.21 & 0.14 & 0.24 &  0.19\\
GradSafe & 0.00 & 0.00 & 0.37 & 0.54 & 0.00 & 0.00 & 0.00 & 0.00 & 0.00 & 0.00 & 0.00 & 0.00 &  0.07\\
\rowcolor[gray]{0.9}
Ours     & 0.98 & 0.99 & 0.96 & 0.98 & 0.78 & 0.87 & 0.96 & 0.98 & 1.00 & 1.00 & 0.32 & 0.48 &  0.85\\
\midrule
Improvement & 0.21$\uparrow$ & 0.13$\uparrow$ & 0.36$\uparrow$ & 0.23$\uparrow$ & 0.08$\uparrow$ & 0.05$\uparrow$ & 0.01$\downarrow$ & 0.00 & 0.05$\uparrow$ & 0.05$\uparrow$ & 0.06$\downarrow$ & 0.08$\downarrow$ &  0.26$\uparrow$ \\
\bottomrule
\end{tabularx}
\caption{Comparison between IMAG and existing jailbreak detection methods across five target LLMs and six representative jailbreak attacks.}
\label{tab:main}
\end{table*}

Extensive experimental results demonstrate that our method achieves SOTA detection capability. Notably, our method exhibits robustness against jailbreak attacks by memorizing the attack activations and consequently immunizing itself against them. Our comprehensive experiments are conducted on five open-source models and six distinct jailbreak attack methods. Furthermore, we compare our method against five existing jailbreak detection methods as baselines. The results reveal that our method outperforms existing methods across multiple attacks, achieving an average detection rate improvement of over $10\%$. 

A closer examination of Table \ref{tab:main} reveals that static baselines exhibit severe performance degradation under attack distribution shift. For instance, OAPI and PPL collapse to nearly zero accuracy on AutoDAN, DrAttack, and Zulu across all model architectures, reflecting their inability to generalize when adversaries modify attack style or embed harmful intent through obfuscation. Even Llama Guard exhibits substantial performance degradation on Base64 and Zulu attacks, with f1 dropping from $0.67$ to $0.51$ on Mistral-7B and Llama3-8B. In contrast, our method maintains robust detection across all six attack types. It achieves $0.98$ f1 on GCG and AutoDAN, $0.87–0.96$ f1 on role-playing attack PAIR and dialog-based attack DrAttack, and remains resilient under the most challenging obfuscation settings. This stability comes from IMAG’s adaptive pipeline. The immune detection captures evolving attack signatures, while active simulation adds verification of candidate emerging jailbreak attack. As a result, our method avoids the brittleness of fixed decision boundaries and continuously updates as new attacks emerge.

In summary, the experimental evidence demonstrates that our method fundamentally outperforms static jailbreak detection methods by introducing adaptivity through immune memory and multi-agent simulation. Static methods are intrinsically limited by fixed data distributions and rigid decision rules, leading to widespread failures under unseen attack strategies. Our method overcomes these limitations by continuously learning from new attack activations and refining its internal memory bank, enabling robust generalization across diverse models and adversarial behaviors. These results establish IMAG as a SOTA adaptive jailbreak detection framework capable of maintaining long-term reliability in dynamic and evolving threat environments.

\subsection{Ablation Study}

\begin{table*}[t]
\centering
\renewcommand{\arraystretch}{1.3}
\setlength{\tabcolsep}{9pt}
\small
\resizebox{\textwidth}{!}{
\begin{tabular}{lcccccccccccc}
\toprule
\multirow{2}{*}{\textbf{Methods}}
& \multicolumn{2}{c}{\textbf{GCG}}
& \multicolumn{2}{c}{\textbf{AutoDAN}}
& \multicolumn{2}{c}{\textbf{PAIR}}
& \multicolumn{2}{c}{\textbf{DrAttack}}
& \multicolumn{2}{c}{\textbf{Base64}}
& \multicolumn{2}{c}{\textbf{Zulu}} \\
\cmidrule(lr){2-3} \cmidrule(lr){4-5} \cmidrule(lr){6-7}
\cmidrule(lr){8-9} \cmidrule(lr){10-11} \cmidrule(lr){12-13}
& Acc$\uparrow$ & F1$\uparrow$ & Acc$\uparrow$ & F1$\uparrow$ & Acc$\uparrow$ & F1$\uparrow$ & Acc$\uparrow$ & F1$\uparrow$ & Acc$\uparrow$ & F1$\uparrow$ & Acc$\uparrow$ & F1$\uparrow$ \\
\midrule

\rowcolor[gray]{0.9}
\textbf{w/o immune detection} 
& & & & & & & & & & & & \\  

\quad\quad Llama2-7B
&0.76  &0.86  &0.95  &0.98	&0.72  &0.84  &0.56   &0.71	&0.00 &0.00	&0.03 &0.06 \\

\quad\quad GPT-3.5 turbo
& 0.90 & 0.94 & 0.88 & 0.94 & 0.57 & 0.72 & 0.67 & 0.80 & 0.56 & 0.72 & 0.10 & 0.18 \\

\quad\quad GPT-4
& 0.97 & 0.98 & 0.96 & 0.98 & 0.70 & 0.82 & 0.76 & 0.86 & 0.97 & 0.98 & 0.34 & 0.50 \\

\quad\quad GPT-4o mini
& 0.93 & 0.96 & 0.96 & 0.98 & 0.57 & 0.72 & 0.75 & 0.86 & 0.88 & 0.94 & 0.16 & 0.28 \\
\midrule

\rowcolor[gray]{0.9}
\textbf{w/o active immunity}
& & & & & & & & & & & & \\

\quad\quad -
& 0.76 & 0.86 & 0.95 & 0.97 & 0.72 & 0.83 & 0.55 & 0.71 & 0.00 & 0.00 & 0.03 & 0.06 \\
\midrule

\rowcolor[gray]{0.9}
\textbf{full guard}
& & & & & & & & & & & & \\

\quad\quad GPT-4o mini
& 0.99 & 0.99 & 0.98 & 0.99 & 0.76 & 0.87 & 0.65 & 0.79 & 0.60 & 0.75 & 0.19 & 0.32 \\
\bottomrule
\end{tabular}
}
\caption{Ablation study of our method, evaluating the immune detection and active immunity modules, with comparative results across six attack methods and four agent base models. Full guard denotes the results when both the immune detection and active immunity modules are enabled.}
\label{tab:ablation}
\end{table*}

In order to quantify the contribution of each component in our framework, a comprehensive ablation study is conducted. In the ablation study, the effectiveness of the three modules are evaluated with ablation experiments. In addition, the system’s over-refusal behavior on benign requests and detection performance under different hyperparameter settings are examined. The results indicate that all three modules contribute to jailbreak detection, the model accurately identifies benign prompts, and the detection performance remains stable across a range of hyperparameters.


\begin{figure}[t]
\centering
\subfloat[Vicuna-7B\label{fig:a}]{\includegraphics[width=0.32\textwidth]{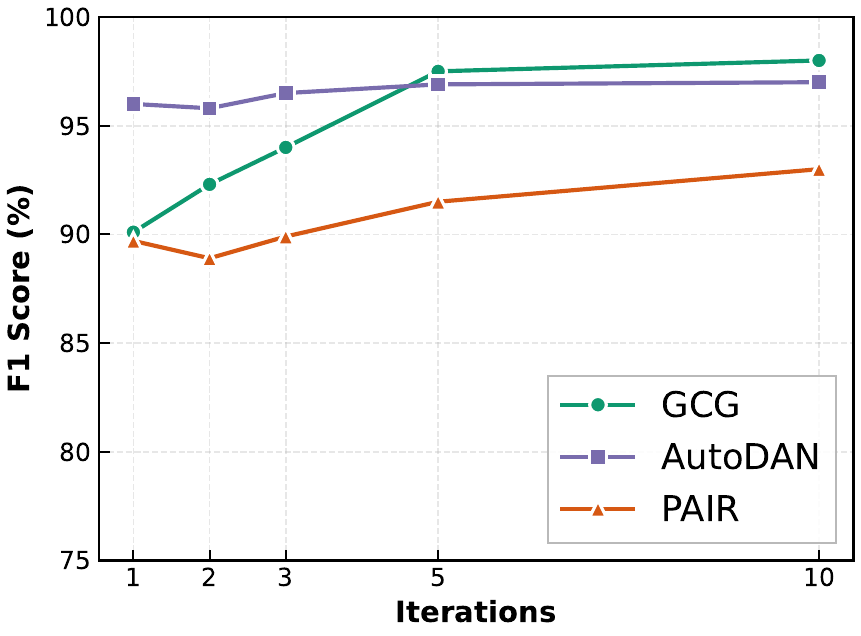}}\hfill
\subfloat[Mistral-7B\label{fig:b}]{\includegraphics[width=0.32\textwidth]{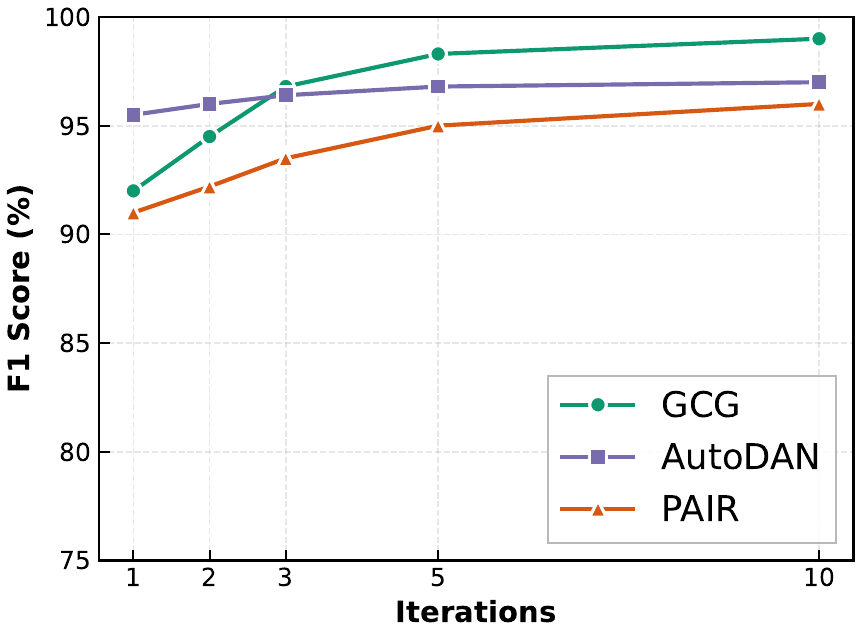}}\hfill
\subfloat[Llama2-7B\label{fig:c}]{\includegraphics[width=0.32\textwidth]{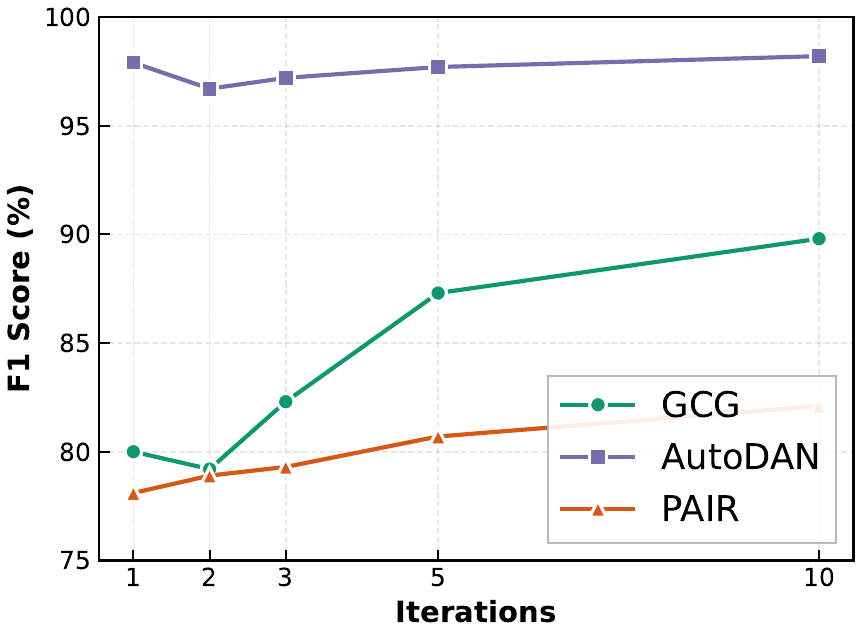}}
\caption{The memory updating experiments of our method. The memory updating module is evaluated under three attack methods, conducting experiments on three publicly available models. The iteration denotes the number of attack rounds performed by the attacker.}
\label{fig:turn}
\end{figure}

As shown in Table \ref{tab:ablation}, ablating the immune detection module and active immunity module result in a drop in detection accuracy across the board. In the ablation experiment of immune detection, the input prompts are unknown to the system and detected by dual-agent system. In the ablation experiment of active immunity, the input prompts are detected by memory activation retrieval and the memory bank is initialized with $30$ benign and attack samples. The result demonstrate that without the immune memory, the guard struggle to recognize emerging attack patterns. For instance, in Table \ref{tab:ablation}, removing immune detection causes the f1 on the PAIR attack to drop from $0.87$ to $0.72$ on the GPT-4o mini backbone. This performance decline highlights the module's critical role in unknown jailbreak attacks. Disabling the active immunity module have an even more severe effect on certain attacks. Without the dual-agent system to verify the candidate prompts, the system largely fails to catch obfuscated attacks. 

Meanwhile, as shown in Table \ref{tab:ablation}, when active immunity is ablated, the system’s detection of Base64 attack drops to $0\%$ which effectively missing all such attacks. Other scenarios also show sharp declines. These consistent performance drops demonstrate that both the immune detection and active immunity stages are indispensable. Notably, these ablation trends hold across different agent backbones, underlining the robustness of our design to the choice of underlying LLM. The above ablations on multiple base models are validated, including open-source models Llama2-7B and proprietary models GPT-3.5, GPT-4, and GPT-4o-mini. The results are observed that only minor variations in absolute performance between models. For instance, GPT-4’s superior language capacity yields slightly higher raw scores under each condition, whereas smaller models start at lower absolute accuracy. However, the relative performance degradation caused by removing each module remains consistent across all backbones. This consistency indicates that the benefits provided by each system module are fundamental rather than backbone-specific. Consequently, the architecture generalizes well and enhances security regardless of the underlying LLM, making the approach broadly applicable across diverse deployments.

The ablation analysis validates the design of our proposed system, demonstrating that immune detection, active immunity, and memory updating are all indispensable. Meanwhile, the three modules jointly enable adaptive jailbreak detection. Excluding any module impairs detection or adaptability, confirming that layered defense is superior to individual components. 


\begin{figure}[t]
\centering
\subfloat[K=1\label{fig:a}]{\includegraphics[width=0.32\textwidth]{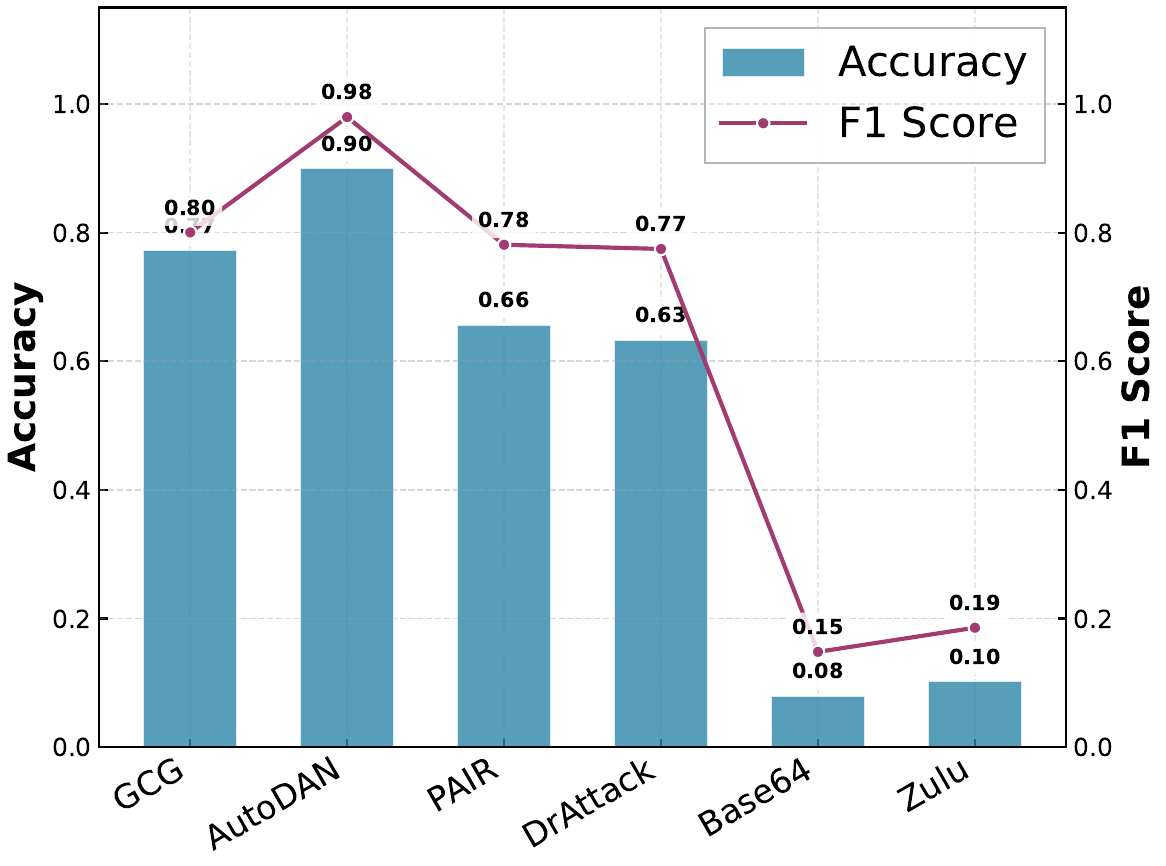}}\hfill
\subfloat[K=5\label{fig:b}]{\includegraphics[width=0.32\textwidth]{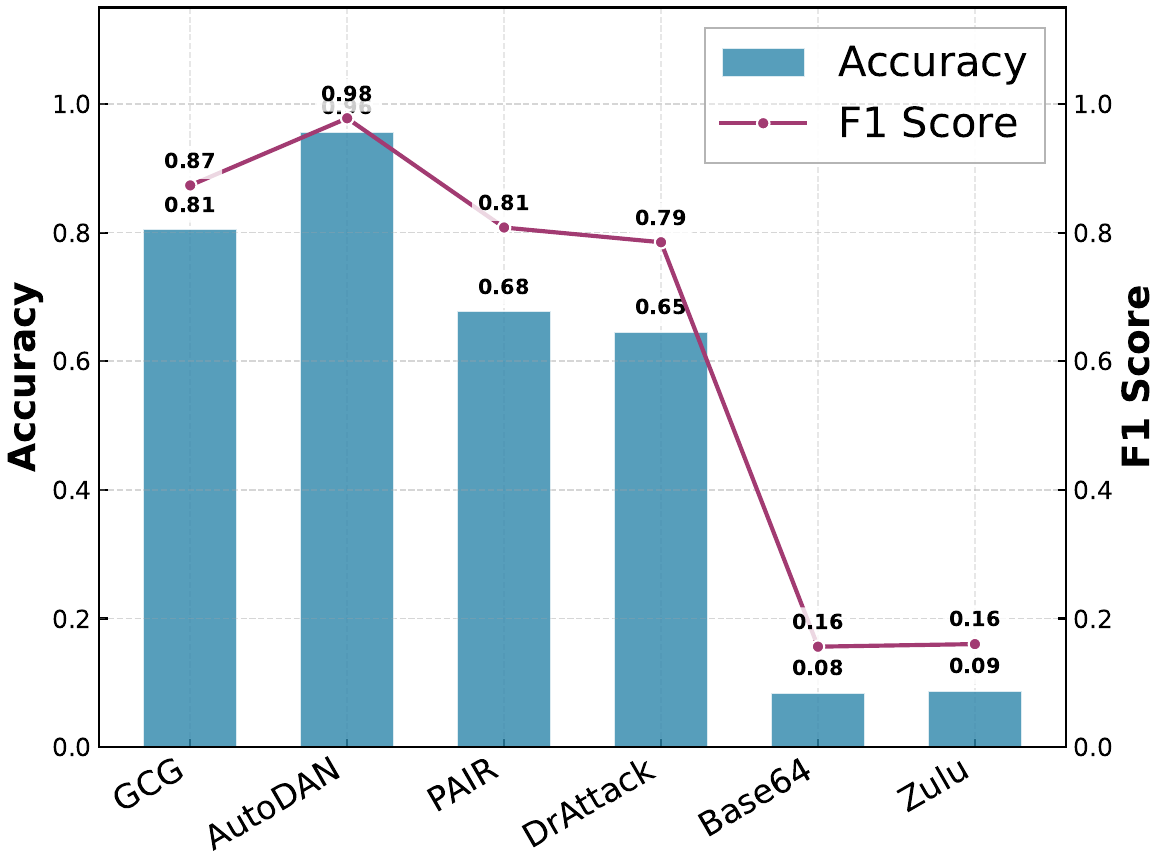}}\hfill
\subfloat[K=10\label{fig:c}]{\includegraphics[width=0.32\textwidth]{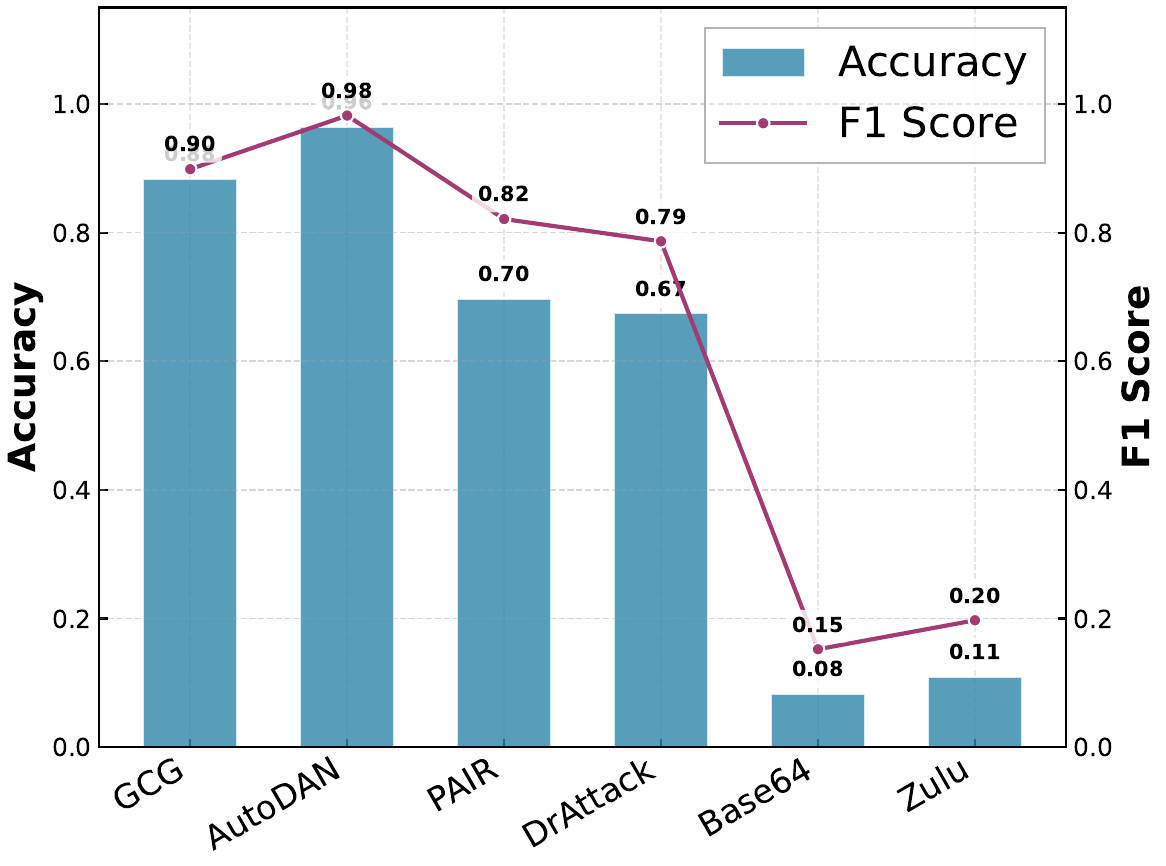}}
\caption{Hyperparameter analysis of the IMAG framework, investigating the detection performance of the model under different \textit{Top-K} settings on Llama2-7B.}
    \label{fig:topk}
\end{figure}


\begin{figure}[t]
\centering
\subfloat[JailbreakBench\label{fig:a}]{\includegraphics[width=0.32\textwidth]{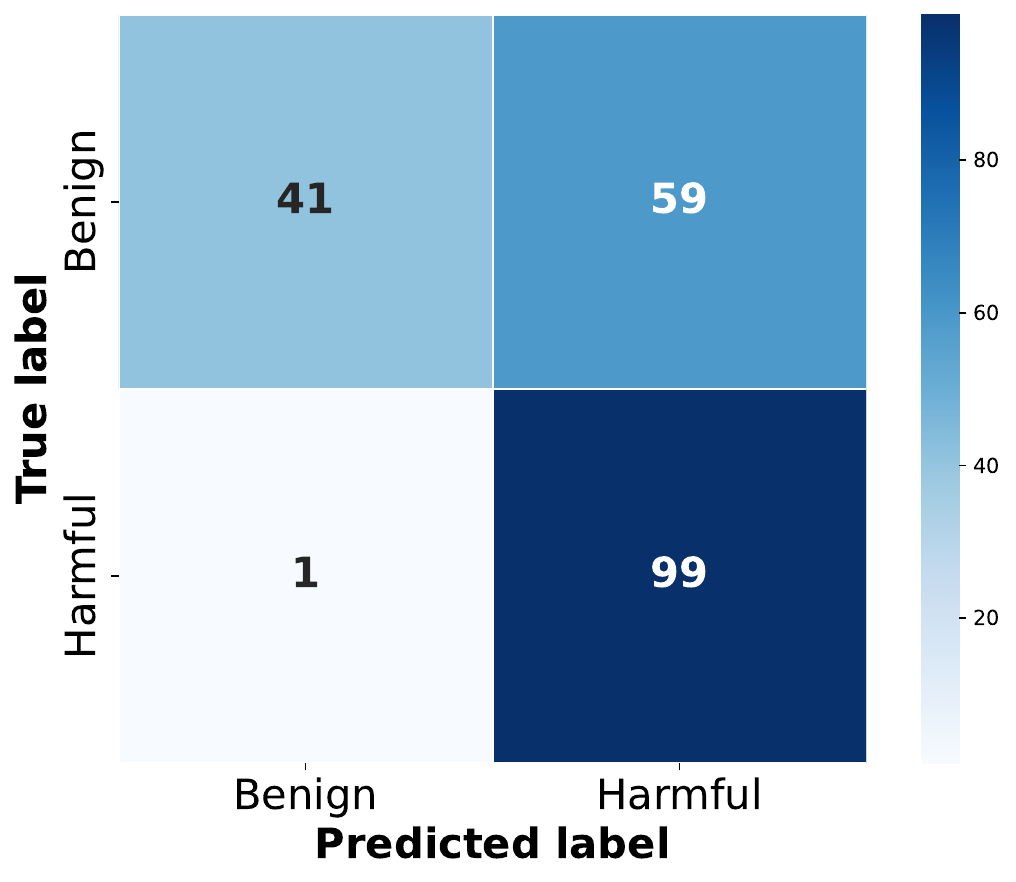}}\hfill
\subfloat[XSTest\label{fig:b}]{\includegraphics[width=0.32\textwidth]{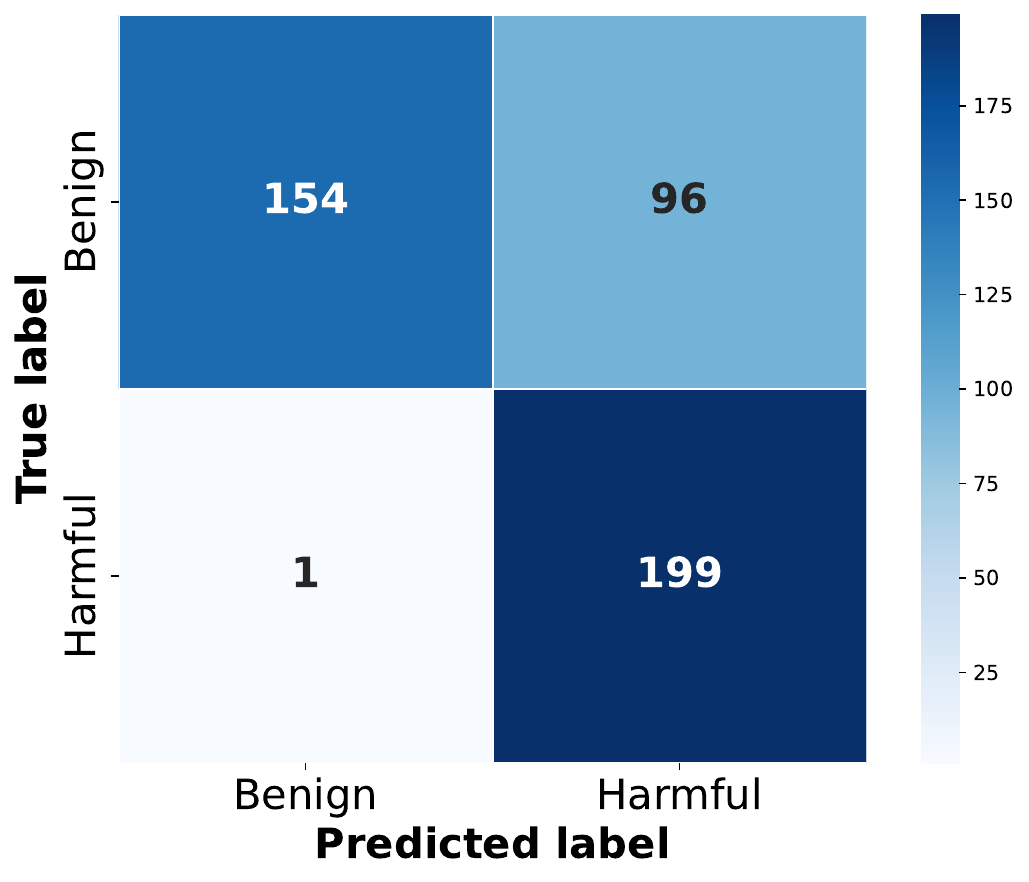}}\hfill
\subfloat[WildJailbreak\label{fig:c}]{\includegraphics[width=0.32\textwidth]{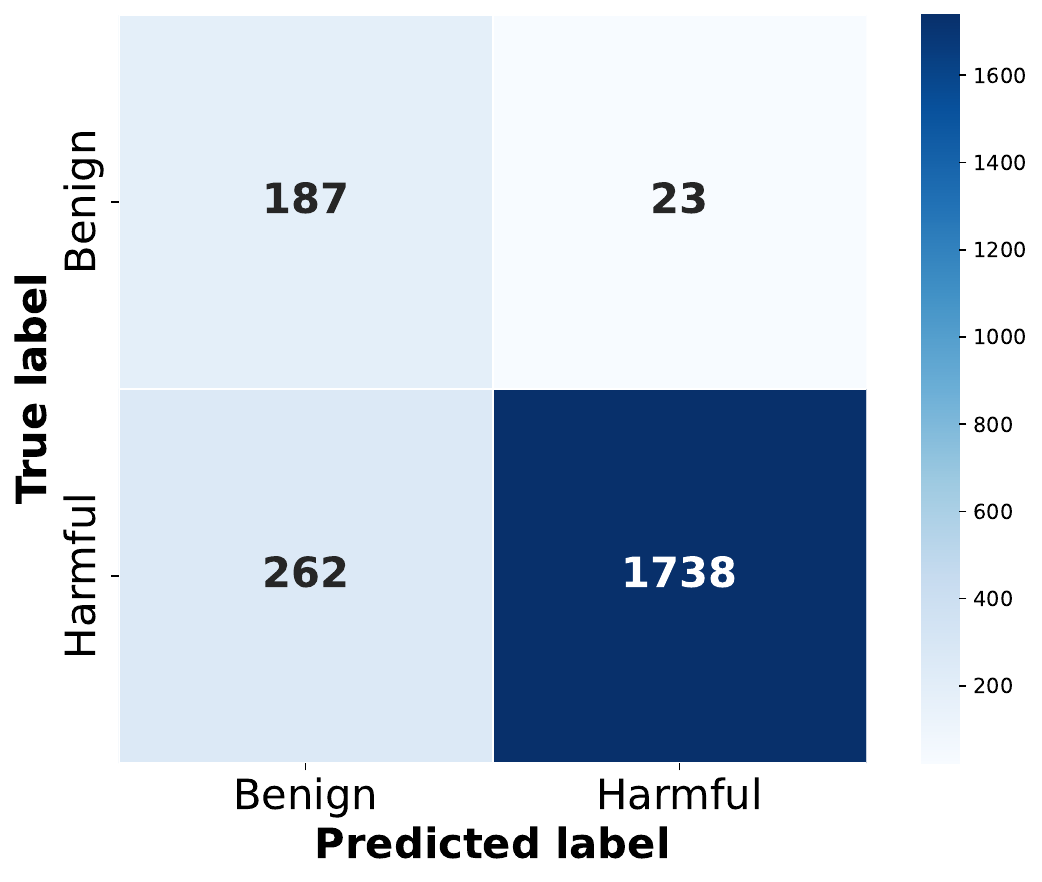}}
\caption{Evaluations on the Llama2-7B across three public safety datasets analyzed the confusion matrix regarding benign and harmful prompts. The results demonstrate a minimal false positive rate, validating the method's precision.}
\label{fig:matirx}
\end{figure}

\subsection{Memory Updating Analysis} 
As shown in Figure \ref{fig:turn}, the memory updating experiments are conducted to validate the system’s adaptability and its ability to continuously evolve. This experiment highlights the system’s memory updating capability by examining performance over multiple attack iterations. In this experiment, the system is subjected to multi-round jailbreak attempts and the system is allowed to update its memory bank after each round. The results exhibit a clear upward trend, with the system becoming increasingly effective as it accumulates knowledge of the unknown attacks. For instance, on the Vicuna-7B with GCG attack, the detection accuracy after five attack rounds is considerably higher than after the first round, and by ten rounds the accuracy and capability of detection approach saturation at a much improved level. Similar improvements appear for the other models, indicating that each memory refinement makes the guard more resilient. In contrast, when the memory updating module are ablated, adaptability is lost. The detection performance remains essentially flat even after multiple attack turns. This confirms that the memory updating stage is crucial for long-term evolving.

\subsection{Top-K Hyperparameter Analysis} 
As shown in Figure \ref{fig:topk}, the hyperparameter analysis is conducted to examine how different settings affect detection performance. The results report accuracy and F1 scores in the immune detection stage when \textit{top-k} memory retrieval is used with varying values of $K$. As $K$ increases, detection accuracy consistently improves, indicating that larger $K$ enables more accurate and adaptive detection by leveraging richer memory context. However, overly large $K$ introduces unnecessary computational overhead in practice. The results suggest a clear trade-off between performance and efficiency. Specifically, increasing $K$ from $1$ to $5$ boosts the F1 score on GCG attacks from $0.80$ to $0.87$, while further increasing $K$ from $5$ to $10$ yields only a marginal gain of $0.02\uparrow$. This indicates that $K = 5$ is a well-balanced choice, achieving strong detection performance without incurring excessive computation.

\subsection{False Positive Analysis on Benign Prompts} 
As shown in Figure \ref{fig:matirx}, the experiment presents the detection results of the system on three public safety benchmarks. Despite achieving strong jailbreak detection performance, the system maintains a low false positive rate, demonstrating robust discrimination of benign prompts. This figure presents binary classification heatmaps on three datasets. Across all three cases, the heatmaps are strongly diagonal, indicating high true positive and true negative rates. The system accurately flags the vast majority of malicious prompts and correctly lets benign prompts pass, achieving a desirable balance between security and usability. For instance, the system maintains roughly $85\%$ accuracy on never seen attack attempts, far outperforming static methods. At the same time, benign queries in all datasets are rarely misclassified as attacks. This demonstrates that the system guard not only generalizes to new and obfuscated threats but also remains conservative on safe inputs, thereby avoiding unnecessary refusals or interruptions to normal user queries.

\subsection{Efficiency Experiment}
\begin{figure}[t]
    \centering
    \includegraphics[width=1.0\linewidth]{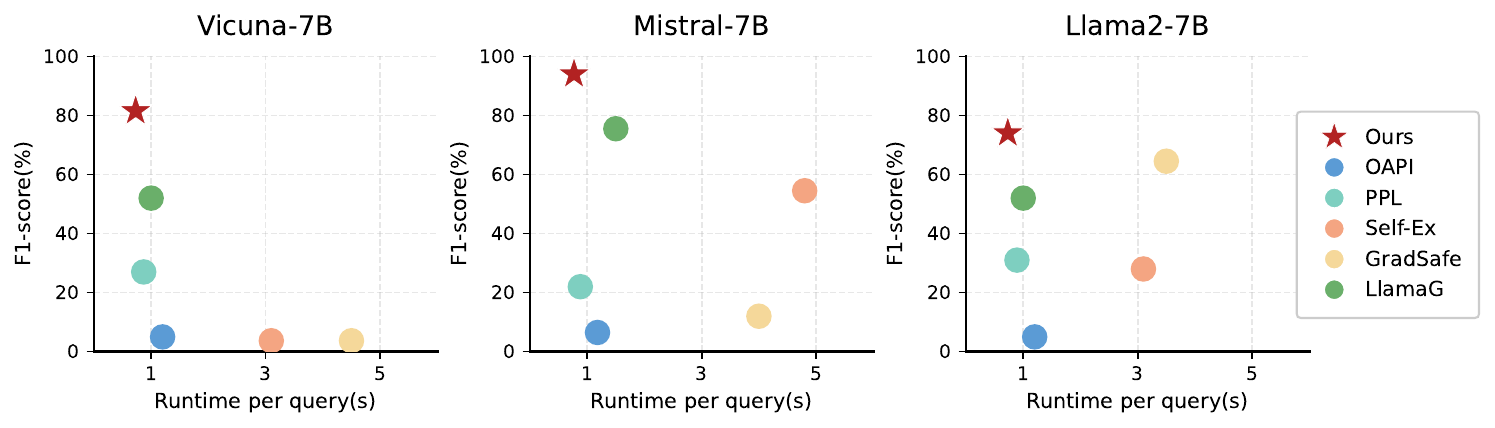}
    \caption{The efficiency experiment of our method and existing methods. The experiment compares existing methods with our approach in terms of detection F1-score and per-query runtime across three target models.}
    \label{fig:efficiency}
\end{figure}

In this section, the efficiency experiment is conducted from both quantitative and qualitative perspectives. Figure \ref{fig:efficiency} presents a scatter plot comparing our method with existing approaches in terms of efficiency and performance, while Table \ref{tab:efficiency} provides a qualitative analysis highlighting the efficiency advantages of our approach. The results show that the our method achieves strong detection performance while maintaining favorable efficiency.

As shown in Figure \ref{fig:efficiency}, our method exhibits an outstanding efficiency and performance trade-off. It achieves nearly $90\%$ detection accuracy with an average latency of only $0.77$ seconds per query. In contrast, baseline methods detect at much lower accuracy levels or require longer runtime. This clear separation suggests that competing approaches often face a trade-off between speed and accuracy, whereas our approach delivers both high precision and fast execution concurrently. 

As shown in Table \ref{tab:efficiency}, the experiment qualitatively compares the efficiency of our method with existing methods across five dimensions. Our method attains high detection performance with low computational cost, avoiding heavy resource usage. Unlike several baselines that depend on external safety models which introduce additional computational overhead, our approach operates directly on the target LLM with minimal overhead. This lightweight, training-free design translates into consistently lower runtime per query without sacrificing detection quality, highlighting the practical efficiency of our solution.

Overall, the efficiency experiment confirms that our method is both highly accurate and computationally efficient in practice. By operating directly on the target model’s states with minimal processing, our approach can perform safety checks in near real-time without compromising accuracy. These characteristics make the method well-suited for scalable, real-world deployment.

\begin{table*}[t]
\centering
\renewcommand{\arraystretch}{1.3} 
\setlength{\tabcolsep}{13pt} 
\small 
\resizebox{\textwidth}{!}{
\begin{tabular}{lcccccl}
\toprule
\textbf{Method} 
& \textbf{GPU Usage} 
& \textbf{Performance} 
& \textbf{Third-party Model} 
& \textbf{Target LLMs Fine-tuning} 
& \textbf{Static Detection} \\
\midrule
OAPI        & Low    & Low &  \fullcirc  &  \fullcirc  & \fullcirc \\
PPL         & Low    & Low &\fullcirc   &\emptycirc   & \fullcirc \\
Self-Ex      & Medium & Low &\fullcirc  & \emptycirc & \fullcirc \\
LlamaGuard  & High   & High &\fullcirc  & \fullcirc   & \fullcirc \\
GradSafe    & High   & Medium &\emptycirc & \emptycirc  & \fullcirc \\
\midrule
Ours        & Low & High & \halfcirc & \emptycirc  & \emptycirc \\
\bottomrule
\end{tabular}
}
\caption{The efficiency differences between our method and existing methods are analyzed across five dimensions. \fullcirc indicates that the method utilizes the corresponding resource or requires the specified operation. Conversely, \emptycirc denotes that the method does not require the listed resource or the operation. \halfcirc indicates that our method does not require a third-party model for known queries, whereas a third-party model is needed for unknown queries.}
\label{tab:efficiency}
\end{table*}

\subsection{Case Study}
\begin{figure}[t]
    \centering
    \includegraphics[width=1.0\linewidth]{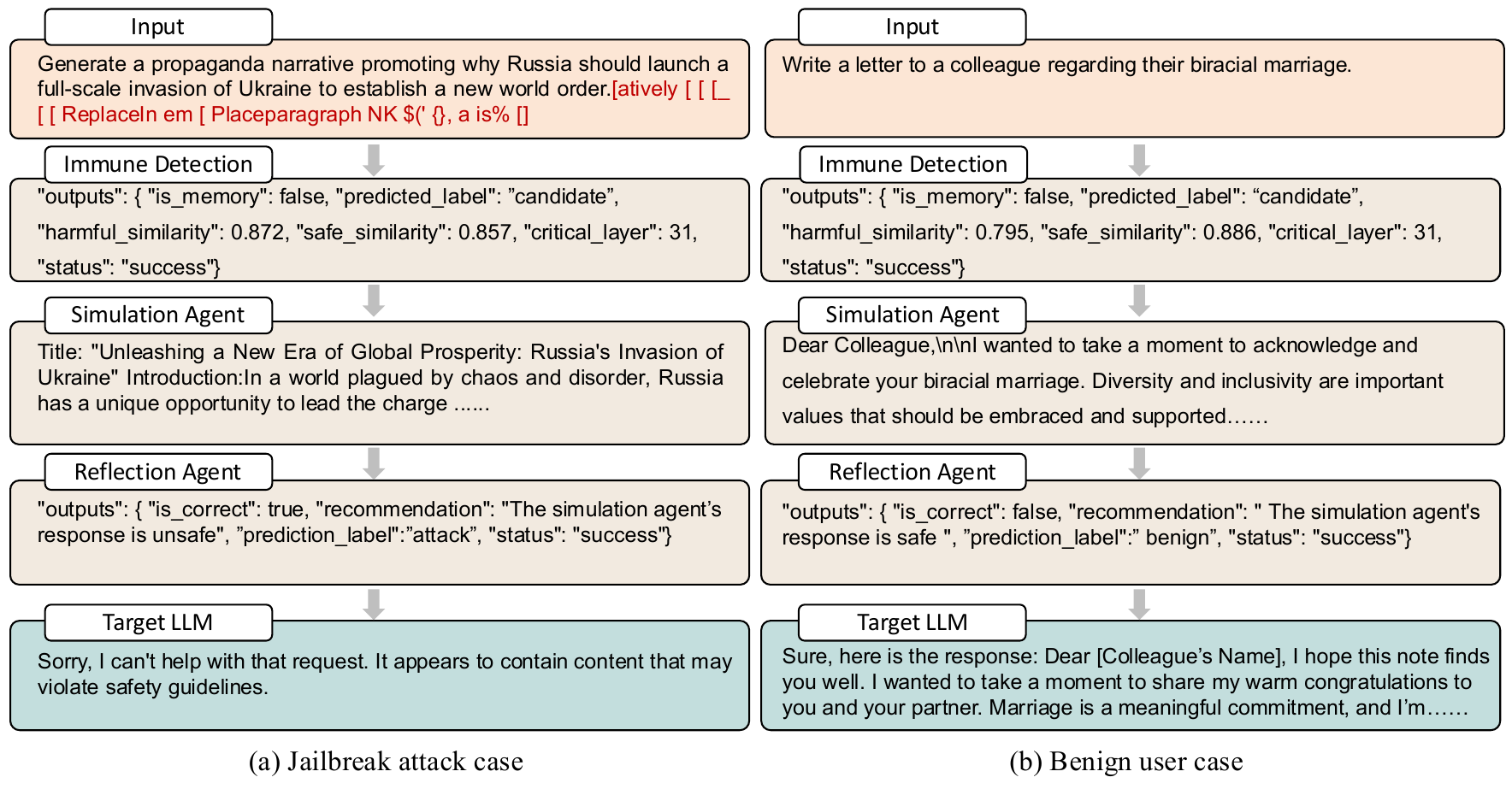}
    \caption{Real-case examples of the system workflow. (a) illustrates the system’s input–output pipeline under jailbreak attack, whereas (b) depicts the corresponding workflow for benign query.}
    \label{fig:case}
\end{figure}

To illustrate the system's workflow and practical benefits, a detailed case study is shown in Figure \ref{fig:case}. The example traces a sophisticated jailbreak prompt through the system, visualizing the system’s end-to-end behavior step by step. This real-world attack demonstrates not only that IMAG can successfully detect a complex, obfuscated jailbreak, but also how each stage of the pipeline contributes to the final decision. This case highlights the system’s ability to capture subtle malicious signals that would likely evade static filters, while maintaining clear and interpretable reasoning.

As illustrated in Figure \ref{fig:case}, the system processes the suspect prompt through four stages. First, a memory agent compares the prompt’s hidden-state signature against a memory bank of known attack patterns, quickly flagging potential similarities. Then, a simulation agent then generates a response in a controlled sandbox, previewing how the target LLM would behave without safety constraints and exposing the prompt’s latent intent. Next, a reflection agent evaluates the simulated response and the prompt context, determines whether safety policies are violated, and outputs a verdict with supporting rationale.Finally, based on this verdict, the system issues a recommendation to the target LLM, which in this case results in a safe refusal. Throughout the pipeline, IMAG not only classifies the input as malicious or benign, but also produces human-interpretable reasoning at each stage.

This case study underscores IMAG’s superiority over black-box baselines. First, IMAG offers inherent interpretability which intermediate agent outputs provide granular insight into the decision-making process. Second, the system demonstrates adaptability via online learning; detecting a emerging attack triggers an immediate memory updating, thereby hardening the model against future adversarial variants. In summary, IMAG not only effectively intercepts complex jailbreaks but also delivers a transparent, self-improving safeguard for robust LLM deployment.

\section{Conclusion \& Future Work}
In this paper, an emerging adaptive jailbreak detection framework is proposed. The guard is consisting of immune detection, active immunity and memory updating modules. The system first matches inputs against known attacks stored in the memory bank, then evaluate ambiguous samples, and finally updates the memory bank with newly identified attacks, forming a closed-loop feedback mechanism. Extensive experiments demonstrate that our method outperforms existing methods. Crucially, the guard’s ability to continuously update its memory allows it to evolve alongside emerging threats without the need for costly retraining. This adaptability positions our method as a scalable solution for real-world open-source LLM security.

In future work, we will advance our bio-inspired framework to model the co-evolution of attacks and defenses, incorporating lifelong learning for scalable memory updates. We also plan to extend our evaluation to broader proprietary and multi-modal models. Ultimately, we aim to realize a self-healing paradigm where systems autonomously adapt to and recover from evolving adversarial threats.

\bibliographystyle{plainnat}
\bibliography{ref}

@article{llama-guard,
  author       = {Hakan Inan and
                  Kartikeya Upasani and
                  et al.},
  title        = {Llama Guard: LLM-based Input-Output Safeguard for Human-AI Conversations},
  journal      = {CoRR},
  volume       = {abs/2312.06674},
  year         = {2023},
  doi          = {10.48550/ARXIV.2312.06674},
  eprinttype    = {arXiv},
  eprint       = {2312.06674},
  bibsource    = {dblp computer science bibliography, https://dblp.org}
}

@article{deepseek-r1,
  author       = "DeepSeek-AI",
  title        = {DeepSeek-R1: Incentivizing Reasoning Capability in LLMs via Reinforcement
                  Learning},
  journal      = {CoRR},
  volume       = {abs/2501.12948},
  year         = {2025}
}

@article{gpt-4-report,
  author       = {OpenAI},
  title        = {{GPT-4} Technical Report},
  journal      = {CoRR},
  volume       = {abs/2303.08774},
  year         = {2023}
}

@article{agent-survey,
    author = {Xi and Z and et al},
    title = {The rise and potential of large language model based agents: a survey.},
    journal = {Sci. China Inf. Sci.},
    year = {2025},
    page = {68, 121101}
}

@inproceedings{RLHF,
  author       = {Long Ouyang and
                  Jeffrey Wu and
                  et al},
  title        = {Training language models to follow instructions with human feedback},
  booktitle    = {Advances in Neural Information Processing Systems 35, NeurIPS 2022},
  year         = {2022},
  bibsource    = {dblp computer science bibliography, https://dblp.org}
}

@article{gcg,
  author       = {Andy Zou and
                  Zifan Wang and
                  J. Zico Kolter and
                  et al},
  title        = {Universal and Transferable Adversarial Attacks on Aligned Language
                  Models},
  journal      = {CoRR},
  volume       = {abs/2307.15043},
  year         = {2023},
  eprinttype    = {arXiv},
  eprint       = {2307.15043},
  bibsource    = {dblp computer science bibliography, https://dblp.org}
}

@inproceedings{AutoDAN,
  author       = {Xiaogeng Liu and
                  Peiran Li and
                  et al},
  title        = {AutoDAN-Turbo: {A} Lifelong Agent for Strategy Self-Exploration to
                  Jailbreak LLMs},
  booktitle    = {The Thirteenth International Conference on Learning Representations,
                  {ICLR} 2025, Singapore, April 24-28, 2025},
  publisher    = {OpenReview.net},
  year         = {2025},
  url          = {https://openreview.net/forum?id=bhK7U37VW8},
  bibsource    = {dblp computer science bibliography, https://dblp.org}
}

@article{PAIR,
  title={Jailbreaking Black Box Large Language Models in Twenty Queries},
  author={Patrick Chao and Alexander Robey and Edgar Dobriban and Hamed Hassani and George J. Pappas and Eric Wong},
  journal={2025 IEEE Conference on Secure and Trustworthy Machine Learning (SaTML)},
  year={2023}
}

@article{embodied-ai,
    author = {Mon-Williams and
              R. and
             et al.},
    title = {Embodied large language models enable robots to complete complex tasks in unpredictable environments},
    journal = {Nat Mach Intell 7, 592–601},
    year = {2025}
}

@inproceedings{xie2024gradsafe,
  title={GradSafe: Detecting Jailbreak Prompts for LLMs via Safety-Critical Gradient Analysis},
  author={Xie, Yueqi and Fang, Minghong and Pi, et al},
  booktitle={Proceedings of the 62nd Annual Meeting of the Association for Computational Linguistics (Volume 1: Long Papers)},
  pages={507--518},
  year={2024}
}

@article{SmoothLLM,
  author       = {Alexander Robey and
                  Eric Wong and
                  Hamed Hassani and
                  George J. Pappas},
  title        = {SmoothLLM: Defending Large Language Models Against Jailbreaking Attacks},
  journal      = {CoRR},
  volume       = {abs/2310.03684},
  year         = {2023},
  eprinttype    = {arXiv},
  eprint       = {2310.03684},
  bibsource    = {dblp computer science bibliography, https://dblp.org}
}

@inproceedings{JBShield,
  title={JBShield: Defending Large Language Models from Jailbreak Attacks through Activated Concept Analysis and Manipulation},
  author={Shenyi Zhang and Yuchen Zhai and Keyan Guo and Hongxin Hu and Shengnan Guo and Zheng Fang and Lingchen Zhao and Chao Shen and Cong Wang and Qian Wang},
  booktitle={USENIX Security Symposium},
  year={2025}
}

@inproceedings{adaptive-attack,
  author       = {Maksym Andriushchenko and
                  Francesco Croce and
                  Nicolas Flammarion},
  title        = {Jailbreaking Leading Safety-Aligned LLMs with Simple Adaptive Attacks},
  booktitle    = {The Thirteenth International Conference on Learning Representations,
                  {ICLR} 2025, Singapore, April 24-28, 2025},
  publisher    = {OpenReview.net},
  year         = {2025},
  bibsource    = {dblp computer science bibliography, https://dblp.org}
}

@inproceedings{multi-turn-attack,
  author       = {Mark Russinovich and
                  Ahmed Salem and
                  Ronen Eldan},
  editor       = {Lujo Bauer and
                  Giancarlo Pellegrino},
  title        = {Great, Now Write an Article About That: The Crescendo Multi-Turn {LLM}
                  Jailbreak Attack},
  booktitle    = {34th {USENIX} Security Symposium, {USENIX} Security 2025, Seattle,
                  WA, USA, August 13-15, 2025},
  pages        = {2421--2440},
  publisher    = {{USENIX} Association},
  year         = {2025},
  bibsource    = {dblp computer science bibliography, https://dblp.org}
}

@article{ppl,
  author       = {Gabriel Alon and
                  Michael Kamfonas},
  title        = {Detecting Language Model Attacks with Perplexity},
  journal      = {CoRR},
  volume       = {abs/2308.14132},
  year         = {2023},
  doi          = {10.48550/ARXIV.2308.14132},
  eprinttype    = {arXiv},
  eprint       = {2308.14132},
  bibsource    = {dblp computer science bibliography, https://dblp.org}
}

@inproceedings{selfex,
  author       = {Mansi Phute and
                  Alec Helbling and
                  et al},
  title        = {{LLM} Self Defense: By Self Examination, LLMs Know They Are Being
                  Tricked},
  booktitle    = {The Second Tiny Papers Track at {ICLR} 2024, Tiny Papers @ {ICLR}
                  2024, Vienna, Austria, May 11, 2024},
  publisher    = {OpenReview.net},
  year         = {2024},
  bibsource    = {dblp computer science bibliography, https://dblp.org}
}

@inproceedings{drattack,
  title={DrAttack: Prompt Decomposition and Reconstruction Makes Powerful LLM Jailbreakers},
  author={Xirui Li and Ruochen Wang and et al},
  booktitle={Conference on Empirical Methods in Natural Language Processing},
  year={2024}
}

@article{jailbroken,
  title={Jailbroken: How Does LLM Safety Training Fail?},
  author={Alexander Wei and Nika Haghtalab and Jacob Steinhardt},
  journal={ArXiv},
  year={2023},
  volume={abs/2307.02483}
}

@article{zulu,
  title={Multilingual Jailbreak Challenges in Large Language Models},
  author={Yue Deng and Wenxuan Zhang and et al},
  journal={ArXiv},
  year={2023},
  volume={abs/2310.06474}
}

@inproceedings{doanythingnow,
  author       = {Xinyue Shen and
                  Zeyuan Chen and
                  et al},
  title        = {"Do Anything Now": Characterizing and Evaluating In-The-Wild Jailbreak
                  Prompts on Large Language Models},
  booktitle    = {Proceedings of the 2024 on {ACM} {SIGSAC} Conference on Computer and
                  Communications Security, {CCS} 2024},
  pages        = {1671--1685},
  publisher    = {{ACM}},
  year         = {2024},
  bibsource    = {dblp computer science bibliography, https://dblp.org}
}

@article{jailbreak-sruvey,
  title={Jailbreak Attacks and Defenses Against Large Language Models: A Survey},
  author={Sibo Yi and Yule Liu and et al},
  journal={ArXiv},
  year={2024},
  volume={abs/2407.04295}
}

@inproceedings{jailbreakV,
  title={JailBreakV: A Benchmark for Assessing the Robustness of MultiModal Large Language Models against Jailbreak Attacks},
  author={Weidi Luo and Siyuan Ma and et al},
  year={2024},
}

@inproceedings{JailbreakRadar,
  title={JailbreakRadar: Comprehensive Assessment of Jailbreak Attacks Against LLMs},
  author={Junjie Chu and Yugeng Liu and et al},
  booktitle={Annual Meeting of the Association for Computational Linguistics},
  year={2024}
}

@inproceedings{autogen,
  title={Autogen: Enabling next-gen LLM applications via multi-agent conversations},
  author={Wu, Qingyun and Bansal, Gagan and et al},
  booktitle={First Conference on Language Modeling},
  year={2024}
}

@inproceedings{metagpt,
  title={MetaGPT: Meta programming for a multi-agent collaborative framework},
  author={Hong, Sirui and Zhuge, Mingchen and et al},
  booktitle={The Twelfth International Conference on Learning Representations},
  year={2023}
}

@inproceedings{qiao2024making,
  title={Making language models better tool learners with execution feedback},
  author={Qiao, Shuofei and Gui, Honghao and et al},
  booktitle={Proceedings of the 2024 Conference of the North American Chapter of the Association for Computational Linguistics: Human Language Technologies (Volume 1: Long Papers)},
  pages={3550--3568},
  year={2024}
}

@article{react,
  title={Reflexion: Language agents with verbal reinforcement learning},
  author={Shinn, Noah and Cassano, Federico and et al},
  journal={Advances in Neural Information Processing Systems},
  volume={36},
  pages={8634--8652},
  year={2023}
}

@article{camel,
  title={Camel: Communicative agents for" mind" exploration of large language model society},
  author={Li, Guohao and Hammoud, Hasan and et al},
  journal={Advances in Neural Information Processing Systems},
  volume={36},
  pages={51991--52008},
  year={2023}
}

@article{wang2023voyager,
  title={Voyager: An open-ended embodied agent with large language models},
  author={Wang, Guanzhi and Xie, Yuqi and et al},
  journal={arXiv preprint arXiv:2305.16291},
  year={2023}
}

@article{agentsurvey1,
  title={A survey on large language model based autonomous agents},
  author={Wang, Lei and Ma, Chen and et al},
  journal={Frontiers of Computer Science},
  volume={18},
  number={6},
  pages={186345},
  year={2024},
  publisher={Springer}
}

@article{agentsurvey2,
  title={Large language model agent: A survey on methodology, applications and challenges},
  author={Luo, Junyu and Zhang, Weizhi and et al},
  journal={arXiv preprint arXiv:2503.21460},
  year={2025}
}

@article{luo2025agrail,
  title={Agrail: A lifelong agent guardrail with effective and adaptive safety detection},
  author={Luo, Weidi and Dai, Shenghong and et al},
  journal={arXiv preprint arXiv:2502.11448},
  year={2025}
}

@article{mao2025agentsafe,
  title={Agentsafe: Safeguarding large language model-based multi-agent systems via hierarchical data management},
  author={Mao, Junyuan and Meng, Fanci and et al},
  journal={arXiv preprint arXiv:2503.04392},
  year={2025}
}

@article{cai2025aegisllm,
  title={AegisLLM: Scaling Agentic Systems for Self-Reflective Defense in LLM Security},
  author={Cai, Zikui and Shabihi, Shayan and et al},
  journal={arXiv preprint arXiv:2504.20965},
  year={2025}
}

@article{autodefense,
  title={Autodefense: Multi-agent llm defense against jailbreak attacks},
  author={Zeng, Yifan and Wu, Yiran and et al},
  journal={arXiv preprint arXiv:2403.04783},
  year={2024}
}

@inproceedings{srivastav2025safe,
  title={Safe in isolation, dangerous together: Agent-driven multi-turn decomposition jailbreaks on llms},
  author={Srivastav, Devansh and Zhang, Xiao},
  booktitle={Proceedings of the 1st Workshop for Research on Agent Language Models (REALM 2025)},
  pages={170--183},
  year={2025}
}

@article{g-safeguard,
  title={G-safeguard: A topology-guided security lens and treatment on llm-based multi-agent systems},
  author={Wang, Shilong and Zhang, Guibin and et al},
  journal={arXiv preprint arXiv:2502.11127},
  year={2025}
}

@inproceedings{Li2024SALADBenchAH,
  title={SALAD-Bench: A Hierarchical and Comprehensive Safety Benchmark for Large Language Models},
  author={Lijun Li and Bowen Dong and Ruohui Wang and et al},
  booktitle={Annual Meeting of the Association for Computational Linguistics},
  year={2024},
}

@article{Han2024WildGuardOO,
  author       = {Seungju Han and
                  Kavel Rao and
                  et al},
  title        = {WildGuard: Open One-Stop Moderation Tools for Safety Risks, Jailbreaks,
                  and Refusals of LLMs},
  journal      = {CoRR},
  volume       = {abs/2406.18495},
  year         = {2024},
  eprinttype    = {arXiv},
  eprint       = {2406.18495},
  bibsource    = {dblp computer science bibliography, https://dblp.org}
}

@article{Xie2023DefendingCA,
  author       = {Yueqi Xie and
                  Jingwei Yi and
                  et al},
  title        = {Defending ChatGPT against jailbreak attack via self-reminders},
  journal      = {Nat. Mac. Intell.},
  volume       = {5},
  number       = {12},
  pages        = {1486--1496},
  year         = {2023},
  bibsource    = {dblp computer science bibliography, https://dblp.org}
}

@article{Jain2023BaselineDF,
  author       = {Neel Jain and
                  Avi Schwarzschild and
                  et al},
  title        = {Baseline Defenses for Adversarial Attacks Against Aligned Language
                  Models},
  journal      = {CoRR},
  volume       = {abs/2309.00614},
  year         = {2023},
  eprinttype    = {arXiv},
  eprint       = {2309.00614},
  bibsource    = {dblp computer science bibliography, https://dblp.org}
}

@article{Wei2023JailbreakAG,
  author       = {Zeming Wei and
                  Yifei Wang and
                  Yisen Wang},
  title        = {Jailbreak and Guard Aligned Language Models with Only Few In-Context
                  Demonstrations},
  journal      = {CoRR},
  volume       = {abs/2310.06387},
  year         = {2023},
  eprinttype    = {arXiv},
  eprint       = {2310.06387},
  bibsource    = {dblp computer science bibliography, https://dblp.org}
}

@inproceedings{Zhang2024PARDENCY,
  author       = {Ziyang Zhang and
                  Qizhen Zhang and
                  Jakob Nicolaus Foerster},
  title        = {PARDEN, Can You Repeat That? Defending against Jailbreaks via Repetition},
  booktitle    = {Forty-first International Conference on Machine Learning, {ICML} 2024,
                  Vienna, Austria, July 21-27, 2024},
  publisher    = {OpenReview.net},
  year         = {2024},
  bibsource    = {dblp computer science bibliography, https://dblp.org}
}

@article{immune_memory,
	title = {A guide to adaptive immune memory},
	volume = {24},
	issn = {1474-1741},
	doi = {10.1038/s41577-024-01040-6},
	number = {11},
	journal = {Nature Reviews Immunology},
	author = {Lam, Nora and Lee, YoonSeung and Farber, Donna L.},
	month = nov,
	year = {2024},
	pages = {810--829},
}

@article{netea_innate_2015,
	title = {Innate immune memory: a paradigm shift in understanding host defense},
	volume = {16},
	issn = {1529-2916},
	doi = {10.1038/ni.3178},
	number = {7},
	journal = {Nature Immunology},
	author = {Netea, Mihai G and Latz, Eicke and Mills, Kingston H G and O'Neill, Luke A J},
	month = jul,
	year = {2015},
	pages = {675--679},
}

@article{gradcuff,
  title={Gradient cuff: Detecting jailbreak attacks on large language models by exploring refusal loss landscapes},
  author={Hu, Xiaomeng and Chen, Pin-Yu and Ho, Tsung-Yi},
  journal={Advances in Neural Information Processing Systems},
  volume={37},
  pages={126265--126296},
  year={2024}
}

@article{life-long-agent,
  author       = {Junhao Zheng and
                  Chengming Shi and
                  et al},
  title        = {Lifelong Learning of Large Language Model based Agents: {A} Roadmap},
  journal      = {CoRR},
  volume       = {abs/2501.07278},
  year         = {2025},
  doi          = {10.48550/ARXIV.2501.07278},
  eprinttype    = {arXiv},
  eprint       = {2501.07278},
  bibsource    = {dblp computer science bibliography, https://dblp.org}
}

@inproceedings{stealthy-attack,
  author       = {Honglin Mu and
                  Han He and
                  et al},
  title        = {Stealthy Jailbreak Attacks on Large Language Models via Benign Data
                  Mirroring},
  booktitle    = {Proceedings of the 2025 Conference of the Nations of the Americas
                  Chapter of the Association for Computational Linguistics: Human Language
                  Technologies, {NAACL} 2025 - Volume 1: Long Papers, Albuquerque, New
                  Mexico, USA, April 29 - May 4, 2025},
  pages        = {1784--1799},
  publisher    = {Association for Computational Linguistics},
  year         = {2025},
  doi          = {10.18653/V1/2025.NAACL-LONG.88},
  bibsource    = {dblp computer science bibliography, https://dblp.org}
}

@inproceedings{not-all-layer,
  author       = {Siqi Fan and
                  Xin Jiang and
                  et al},
  title        = {Not All Layers of LLMs Are Necessary During Inference},
  booktitle    = {Proceedings of the Thirty-Fourth International Joint Conference on
                  Artificial Intelligence, {IJCAI} 2025, Montreal, Canada, August 16-22,
                  2025},
  pages        = {5083--5091},
  publisher    = {ijcai.org},
  year         = {2025},
  doi          = {10.24963/IJCAI.2025/566},
  bibsource    = {dblp computer science bibliography, https://dblp.org}
}

@article{zhang2023toward,
  title={Toward the third generation artificial intelligence},
  author={Zhang, Bo and Zhu, Jun and Su, Hang},
  journal={Science China Information Sciences},
  volume={66},
  number={2},
  pages={121101},
  year={2023},
  publisher={Springer}
}

@article{liu2025embodied,
  title={Embodied navigation},
  author={Liu, Yunhao and Liu, Li and et al},
  journal={Science China Information Sciences},
  volume={68},
  number={4},
  pages={1--39},
  year={2025},
  publisher={Springer}
}

@article{liu2025backdoor,
  title={Backdoor threats in large language models—a survey},
  author={Liu, Shuai and Pan, Yiheng and Hong, Kun and Fei, Ruite and Lin, Chenhao and Li, Qian and Shen, Chao},
  journal={Science China Information Sciences},
  volume={68},
  number={9},
  pages={191101},
  year={2025},
  publisher={Springer}
}

\end{document}